\RequirePackage{fix-cm}

\documentclass[smallextended]{svjour3}       %
\smartqed  

\usepackage{graphicx}
\usepackage{nicefrac}
\usepackage[T1]{fontenc}
\usepackage{xspace}
\usepackage{flushend}
\usepackage{color}
\usepackage{url}
\usepackage{multirow}
\usepackage{listliketab}
\usepackage{amssymb}
\usepackage{algorithm}
\usepackage[misc]{ifsym}
\usepackage{algorithmic}
\usepackage{booktabs}
\usepackage{comment}
\usepackage{amsmath}
\usepackage{bbding}
\usepackage{listings}
\usepackage{float}
\usepackage{tabularx}
\usepackage{cite}
\usepackage{rotating}
\usepackage[table,xcdraw]{xcolor}
\usepackage{natbib}
\usepackage{tcolorbox}
\usepackage{ulem}
\usepackage{censor}
\usepackage{tikz}
\usepackage{amssymb} 
\usepackage{amsfonts}
\usepackage{graphicx,subfigure}
\usepackage{xparse}
\usepackage[colorinlistoftodos]{todonotes}
\usepackage{ulem}
\usepackage[hyphenbreaks]{breakurl}
\usepackage{wrapfig}

\usetikzlibrary{arrows,shapes,positioning}
\usetikzlibrary{decorations.markings}
\tikzstyle arrowstyle=[scale=1]
\tikzset{>=latex}

\usepackage[colorlinks = false,
            linkcolor = blue,
            urlcolor  = blue,
            citecolor = blue,
            anchorcolor = blue]{hyperref}

\newlength\CMAX  

\newcommand\PqOne[0]{$PQ_1$}
\newcommand\PqTwo[0]{$PQ_2$}
\newcommand\RqOne[0]{$RQ_1$}
\newcommand\RqTwo[0]{$RQ_2$}

\newcommand{\llags}{Lags}
\newcommand{\lags}{lags}

\newcommand{\lag}{lag}
\newcommand{\snyk}{snyk.io}
\newcommand{\fixupdate}{package-side fixing release}
\newcommand{\ffixupdate}{Package-side fixing release}
\newcommand{\fixupdatetype}{package landing}
\newcommand{\fixupdatetypeS}[1]{package #1 landing}
\newcommand{\ffixupdatetypeS}[1]{Package #1 landing}

\newcommand{\vulfix}{fixing}
\newcommand{\clientupdate}{client-side fixing release}
\newcommand{\clientupdatetype}{client landing}
\newcommand{\clientupdatetypeS}[1]{client #1 landing}
\newcommand{\cclientupdatetypeS}[1]{Client #1 landing}
\newcommand{\cclientupdate}{Client-side fixing release}

\newcommand{\PqOneSen}{(\PqOne) What is the prevalence of \fixupdatetypeS{patch}?}
\newcommand{\PqTwoSen}{(\PqTwo) What portion of the release content is a vulnerability fix?}

\newcommand{\RqOneSen}{(\RqOne) Is the \fixupdate~consistent with the \clientupdate?}
\newcommand{\RqTwoSen}{(\RqTwo) Do lineage freshness and severity influence lags in the fix propagation?}

\colorlet{punct}{red!60!black}
\definecolor{background}{HTML}{EEEEEE}
\definecolor{delim}{RGB}{20,105,176}
\colorlet{numb}{magenta!60!black}

\lstdefinelanguage{json}{
    basicstyle=\small,
    numbers=left,
    numberstyle=\scriptsize,
    stepnumber=1,
    numbersep=8pt,
    showstringspaces=false,
    breaklines=true,
    frame=lines,
    backgroundcolor=\color{background},
    literate=
     *{0}{{{\color{numb}0}}}{1}
      {1}{{{\color{numb}1}}}{1}
      {2}{{{\color{numb}2}}}{1}
      {3}{{{\color{numb}3}}}{1}
      {4}{{{\color{numb}4}}}{1}
      {5}{{{\color{numb}5}}}{1}
      {6}{{{\color{numb}6}}}{1}
      {7}{{{\color{numb}7}}}{1}
      {8}{{{\color{numb}8}}}{1}
      {9}{{{\color{numb}9}}}{1}
      {:}{{{\color{punct}{:}}}}{1}
      {,}{{{\color{punct}{,}}}}{1}
      {\{}{{{\color{delim}{\{}}}}{1}
      {\}}{{{\color{delim}{\}}}}}{1}
      {[}{{{\color{delim}{[}}}}{1}
      {]}{{{\color{delim}{]}}}}{1},
}

\begin{document}

\title{\llags~in the Release, Adoption, and Propagation of \texttt{npm} Vulnerability Fixes}

\author{Bodin Chinthanet\and
        Raula Gaikovina Kula\and
        Shane McIntosh\and
        Takashi Ishio\and
        Akinori Ihara\and
        Kenichi Matsumoto%
}

\institute{
    Bodin Chinthanet \Letter, 
    Raula Gaikovina Kula, Takashi Ishio, Kenichi Matsumoto 
    \at Nara Institute of Science and Technology, Japan\\
    \email{\{bodin.chinthanet.ay1,raula-k,ishio,matumoto\}@is.naist.jp}           
    \and
    Shane McIntosh
    \at  University of Waterloo, Canada\\
    \email{shane.mcintosh@uwaterloo.ca}
    \and
    Akinori Ihara 
    \at Wakayama University, Japan\\
    \email{ihara@wakayama-u.ac.jp}       
}

\date{Received: date / Accepted: date}

\maketitle

\begin{abstract}
Security vulnerability in third-party dependencies is a growing concern not only for developers of the affected software, but for the risks it poses to an entire software ecosystem, e.g., Heartbleed vulnerability.
Recent studies show that developers are slow to respond to the threat of vulnerability, sometimes taking four to eleven months to act. 
To ensure quick adoption and propagation of a release that contains the fix (\textit{\vulfix~release}), we conduct an empirical investigation to identify \lags~that may occur between the vulnerable release and its \vulfix~release (\textit{\fixupdate}).
Through a preliminary study of 231 \fixupdate~of npm projects on GitHub, we observe that a \vulfix~release is rarely released on its own, with up to 85.72\% of the bundled commits being unrelated to a fix.
We then compare the \fixupdate~with changes on a client-side (\textit{\clientupdate}). 
Through an empirical study of the adoption and propagation tendencies of 1,290 \fixupdate s that impact throughout a network of 1,553,325 releases of npm packages, we find that stale clients require additional migration effort, even if the \fixupdate~was quick (i.e., \fixupdatetypeS{patch}).
Furthermore, we show the influence of factors such as the branch that the \fixupdate~lands on and the severity of vulnerability on its propagation.
In addition to these \lags~we identify and characterize, this paper lays the groundwork for future research on how to mitigate propagation \lags~in an ecosystem.
\end{abstract}

\section{Introduction}
\label{sec:introduction}

Vulnerability in third-party dependencies is a growing concern for the software developer.
In a 2018 report, over \textit{four million vulnerabilities} were raised to the attention of developers of over 500 thousand GitHub repositories \citep{Web:github_num_vul}.
The risk of vulnerabilities is not restricted to the direct users of these software artifacts, but it also extends to the broader software ecosystems to which they belong. 
Examples include the ShellShock \citep{Shellsho67:online} and Heartbleed \citep{Heartble2:online} vulnerabilities, which caused widespread damages to broad and diverse software ecosystems made up of direct and indirect adopters. 
Indeed, the case of Heartbleed emphasized its critical role in the modern web.
\citet{Durumeric:2014} shows that OpenSSL, i.e., the project where the Heartbleed vulnerability originated, is presented on web servers that host (at least) 66\% of sites and (at least) 24\% of the secure sites on the internet were affected by Heartbleed.

The speed at which ecosystems react to vulnerabilities and the availability of fixes to vulnerabilities is of paramount importance.
Three lines of prior works support this intuition:
\begin{enumerate}
\item Studies by \citet{Howard:2001, Ponta:icsme2018, Nguyen:2016, Munaiah:2017, Hejderup2015, Pashchenko:2018, Williams:2018} encourage developers to use security best practices, e.g., project validation, security monitoring, to prevent and detect vulnerabilities in deployed projects.
\item Studies by \citet{Kikas:2017, Decan:2017, Cox-ICSE2015} show that vulnerabilities can cascade transitively through the package dependency network.
Moreover, they observe that security issues are more likely to occur in the field due to stale (outdated) dependencies than directly within product codes.
\item Studies by \citet{Kula:2017, Bavota:2015, Bogart:2016, Ihara_OSS2017} show that developers are slow to update their vulnerable packages, which is occasionally due to management and process factors.
\end{enumerate}

While these prior studies have made important advances, they have tended to focus on (i) a coarse granularity, i.e., releases, that focused only on the vulnerable dependency and only the direct client.
For example, \citet{Decan:2018} analyzed how and when package releases with vulnerabilities are discovered and fixed with a single direct client, while \citet{Decan:ICSME:2018} analyzed releases to explore the evolution of technical lag and its impact.
Commit-level analysis similar to \cite{Li:2017, Piantadosi:2019} is important because it reveals how much development activity (i.e., migration effort) is directed towards fixing vulnerabilities compare to the other tasks.
Furthermore, there is also a research gap that relates to (ii) the analysis of the package vulnerability fixes with respect to the downstream clients, and not a single direct client.

To bridge these two research gaps, we set out to identify and characterize the release, adoption, and propagation tendencies of vulnerability fixes.
We identify and track a release that contains the fix, which is defined as a \textit{\vulfix~release}.
We then characterize the \vulfix~release for each npm JavaScript package in terms of commits for fixing the vulnerability, which is defined as the \textit{\fixupdate}.
Based on semantic versioning \citep{Web:semver}, a \fixupdate~is a \texttt{\fixupdatetypeS{major}}, \texttt{\fixupdatetypeS{minor}}, or \texttt{\fixupdatetypeS{patch}}.
From a client perspective, we identify \textit{\clientupdate} lags to classify how a client migrates from a vulnerable version to the \vulfix~version as a \texttt{\clientupdatetypeS{major}}, \texttt{\clientupdatetypeS{minor}}, \texttt{\clientupdatetypeS{patch}}, or \texttt{dependency removal}. 
By comparing the \fixupdate~with the \clientupdate, we identify lags in the adoption process as clients keep stale dependencies.

Our empirical study is comprised of two parts.
First, we perform a preliminary study of 231 \fixupdate s of npm packages on GitHub.
We find that the \fixupdate~is rarely released on its own, with up to 85.72\% of the bundled commits being unrelated to the fix.
Second, we conduct an empirical study of 1,290 \fixupdate s to analyze their adoption and propagation tendencies throughout a network of 1,553,325 releases of npm packages.
We find that quickly releasing fixes does not ensure that clients will adopt them quickly.
Indeed, we find that only 21.28\% of clients reacted to this by performing a \clientupdatetypeS{patch}~of their own.
Furthermore, we find that factors such as the branch upon which a fix lands and the severity of the vulnerability have a small effect on its propagation trajectory throughout the ecosystem, i.e., the latest lineage and medium severity suffer the most \lags.
To mitigate propagation \lags~in an ecosystem, we recommend developers and researchers to (i) develop strategies for making the most efficient update via the release cycle, (ii) develop better awareness mechanisms for quicker planning of an update, and (iii) allocate additional time before updating dependencies.

Our contributions are three-fold.
The first contribution is a set of definitions and measures to characterize the vulnerability discovery and \vulfix~process from both the vulnerable package, i.e., \fixupdate, and its client, i.e., \clientupdate.
The second contribution is an empirical study that identified potential \lags~in the release, adoption, and propagation of a \fixupdate.
\sloppy
The third contribution is a detailed replication package, which is available at \url{https://github.com/NAIST-SE/Vulnerability-Fix-Lags-Release-Adoption-Propagation}.

\subsection{Paper Organization}
The remainder of the paper is organized as follows: 
Section \ref{sec:prelim} describes key concepts and definitions.
Section \ref{sec:prelim_study} presents the motivations, approaches, and results of our preliminary study.
Section \ref{sec:adoption_propagation} introduces the concepts of fix adoption and propagation lags modeling and tracking with the empirical study to identify them.
We then discuss implications of our results and threats to validity of the study in Section \ref{sec:discussion}.
Section \ref{sec:related_work} surveys the related works.
Finally, Section \ref{sec:conclusion} concludes the study.
\section{Concepts and Definitions}
\label{sec:prelim}
\begin{figure*}[hbt]
\centering
\includegraphics[width=1\textwidth]{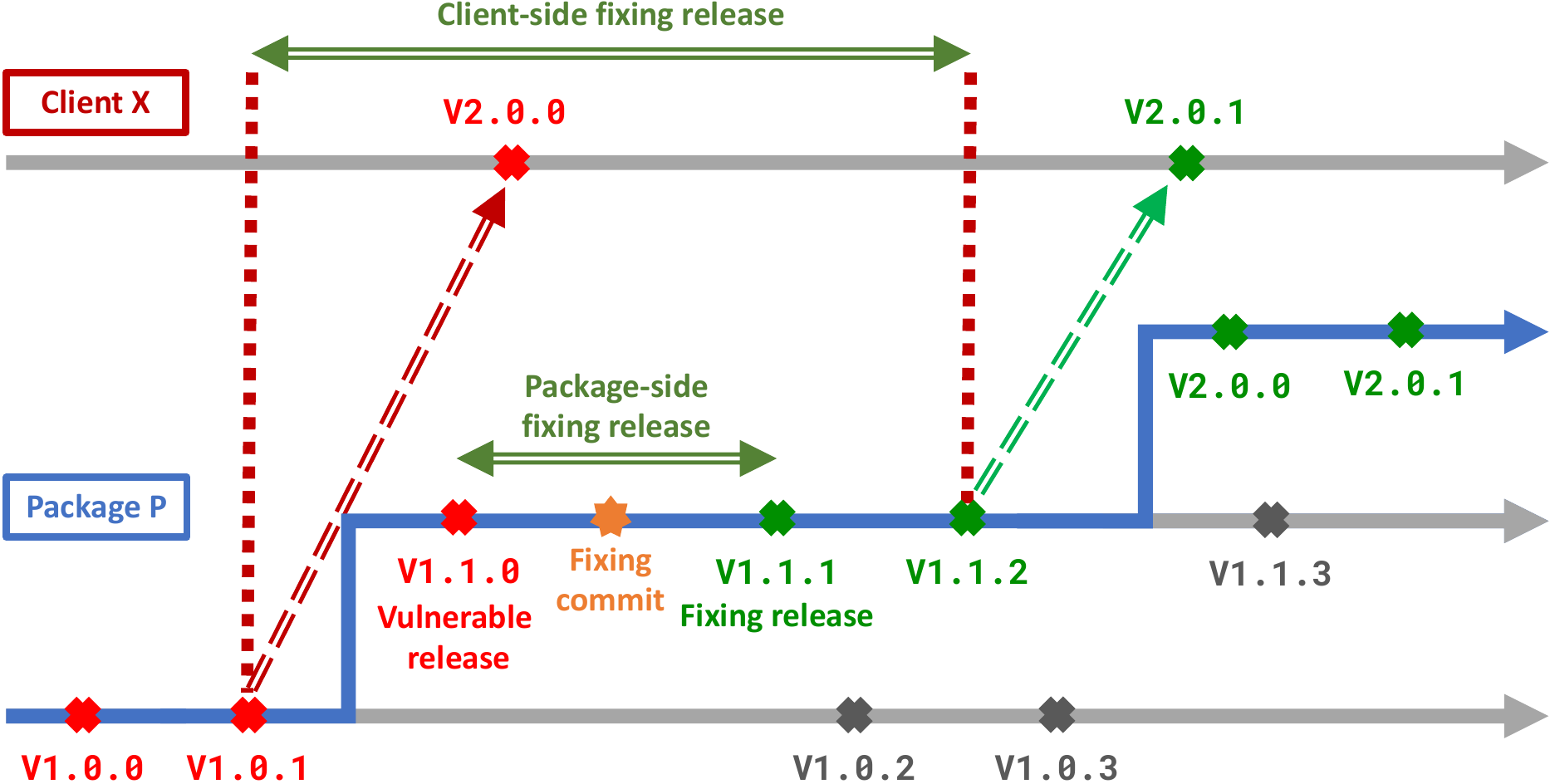}
\caption{The relationship between package-side and client-side regarding vulnerability discovery, \vulfix, and release process of package $\mathbb{P}$ and client $\mathbb{X}$ over time. Red and green releases indicate whether releases are vulnerable or not.}
\label{fig:vul_discovery_n_fix_releasing}
\label{fig:fix_adoption_n_propagation}
\end{figure*}
\subsection{Package-side Vulnerability Fixing Process}
\label{subsec:package-side_fixing_process}

Figure \ref{fig:vul_discovery_n_fix_releasing} illustrates the timeline of the package-side vulnerability fixing process of package $\mathbb{P}$ in the lower part of the figure.
We break down this process into two steps:

    \noindent\uline{Step one: Vulnerability Discovery.} 
    Figure \ref{fig:vul_discovery_n_fix_releasing} shows the vulnerability of package $\mathbb{P}$ being detected after the release of $\mathbb{P}_{V1.1.0}$.
    As reported by \citet{Kula:2017}, CVE defines four phases of a vulnerability: (i) threat detection, (ii) CVE assessment, (iii) security advisory, and (iv) patch release. 
    We define the vulnerability discovery as the period between the threat detection and before the patch release.
    It is most likely that the fixing process starts after the CVE assessment, i.e., a vulnerability has been assigned a CVE number. 
    In addition to a CVE number, a vulnerability report details the affected packages, which can include releases up to an upper-bound of reported versions.
    In this step, the developers of an affected package are notified via communication channels such as a GitHub issue for GitHub projects.
    
    \noindent\uline{{Step two: Vulnerability Fix and Release.}}
    Figure \ref{fig:vul_discovery_n_fix_releasing} shows the vulnerability of package $\mathbb{P}$ being fixed and released as  $\mathbb{P}_{V1.1.1}$. 
    We define the vulnerability fix and release as the period where developers spend their efforts to identify and mitigate the vulnerable code. 
    Once the fix is ready, developers merge that fix to a package repository.
    In most GitHub projects, developers will review changes via a GitHub pull request.
    Semantic versioning convention is also used to manage the release version number of a package \citep{Web:semver}.
    We define the \textit{\vulfix~release} as the first release that contains a vulnerability fix ($\mathbb{P}_{V1.1.1}$). 
    We also define \textit{\fixupdate} to describe the \vulfix~release of the vulnerable package and to show how a package bumped the release version number from a vulnerable release to a \vulfix~release.
    There are three kinds of \fixupdate~based on the semantic versioning: 
    (i) \texttt{\fixupdatetypeS{major}}~(major number is bumped up),
    (ii) \texttt{\fixupdatetypeS{minor}}~(minor number is bumped up), and
    (iii) \texttt{\fixupdatetypeS{patch}}~(patch number is bumped up).
    As shown in Figure \ref{fig:vul_discovery_n_fix_releasing}, \fixupdate~of package $\mathbb{P}$~is \fixupdatetypeS{patch}, i.e., $\mathbb{P}_{V1.1.0}$ to $\mathbb{P}_{V1.1.1}$. 
    
\subsection{Client-side Fixing Release} 
Prior work suggests that lags in adoption could be the result of migration effort \citep{Kula:2017}.
Thus to quantify this effort, we compare a vulnerable release that a client is using, i.e., listed in the report, against the \fixupdate~to categorize as a \textit{\clientupdate}.
Developers of the vulnerable package fulfilled their responsibility, thus the adoption responsibility is left to the client.

Figure \ref{fig:fix_adoption_n_propagation} shows the timelines of a \vulfix~release and its clients. 
As illustrated in the figure, client $\mathbb{X}$ suffers a \lag~in the adoption of a \fixupdate~by switching dependency branches, i.e., \clientupdatetypeS{minor}. 
Also, client $\mathbb{X}$ directly depends on package $\mathbb{P}$.
Client $\mathbb{X}$ is vulnerable (V2.0.0) due to its dependency ($\mathbb{P}_{V1.0.1}$).
To mitigate the vulnerability, package $\mathbb{P}$ creates a new branch, i.e., minor branch, which includes the \fixupdate~($\mathbb{P}_{V1.1.1}$).
Client $\mathbb{X}$ finally adopts the new release of package $\mathbb{P}$ which is not vulnerable (V1.1.2).
We consider that client $\mathbb{X}$ has a \lag, which was not efficient because it actually skipped the \fixupdate~($\mathbb{P}_{V1.1.1}$). 
Instead, client $\mathbb{X}$ adopted the next release ($\mathbb{P}_{V1.1.2}$). 
A possible cause of \lags~is the potential migration effort needed to switch branches, i.e., from $\mathbb{P}_{V1.0.1}$ to $\mathbb{P}_{V1.1.2}$.
The migration effort for major or minor changes may include breaking changes or issues in the release cycle.

\begin{figure*}[]
    \centering
    \subfigure[\texttt{socket.io} vulnerability report with medium severity.]
    {
        \includegraphics[width=0.8\linewidth]{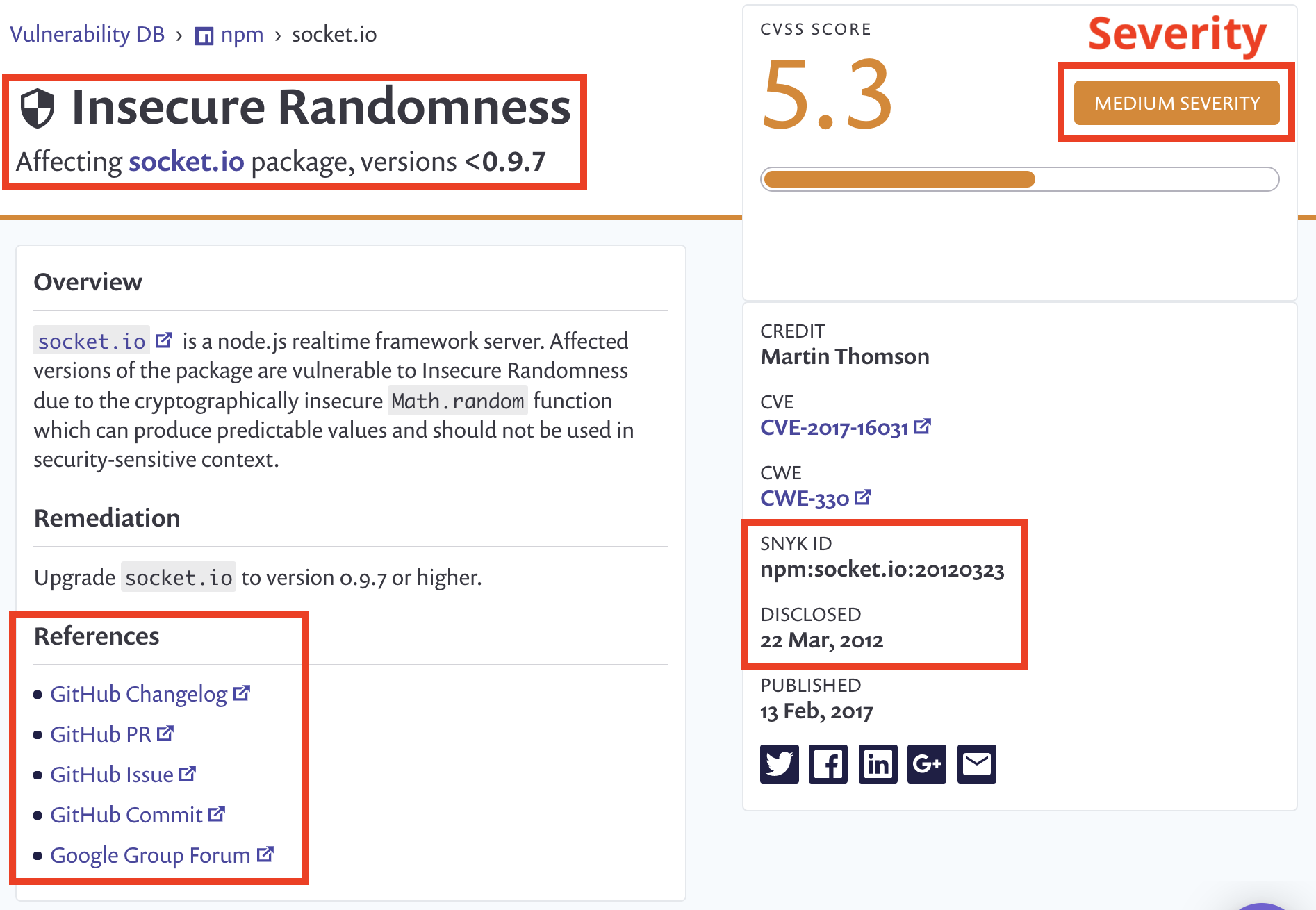}
        \label{fig:vul_report}
    }
    \subfigure[GitHub Issue reporting the vulnerability.]
    {
        \includegraphics[width=0.45\linewidth]{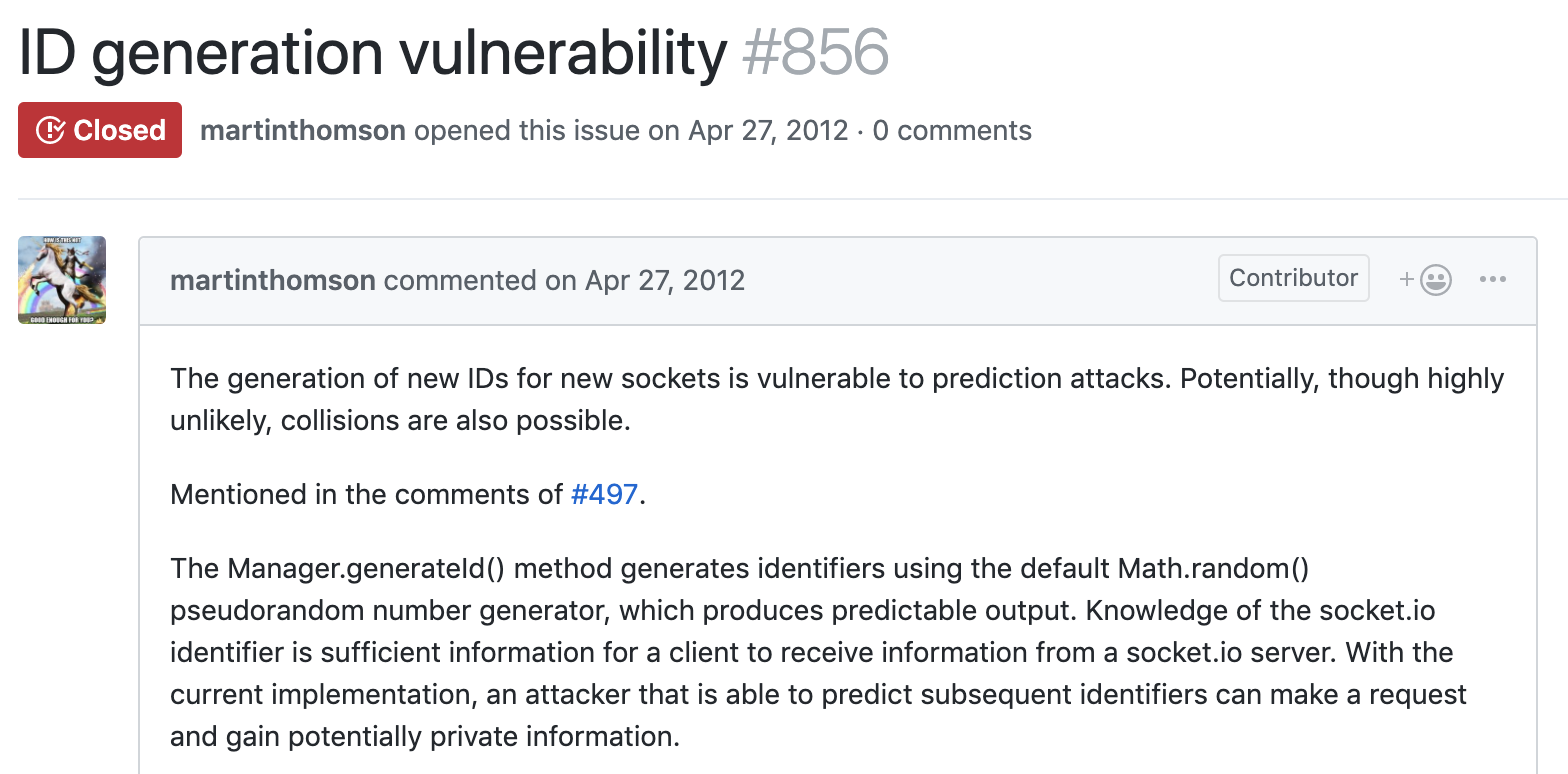}
        \label{fig:github_issue}
    }
    \subfigure[Pull request containing the \vulfix~commit.]
    {
        \includegraphics[width=0.45\linewidth]{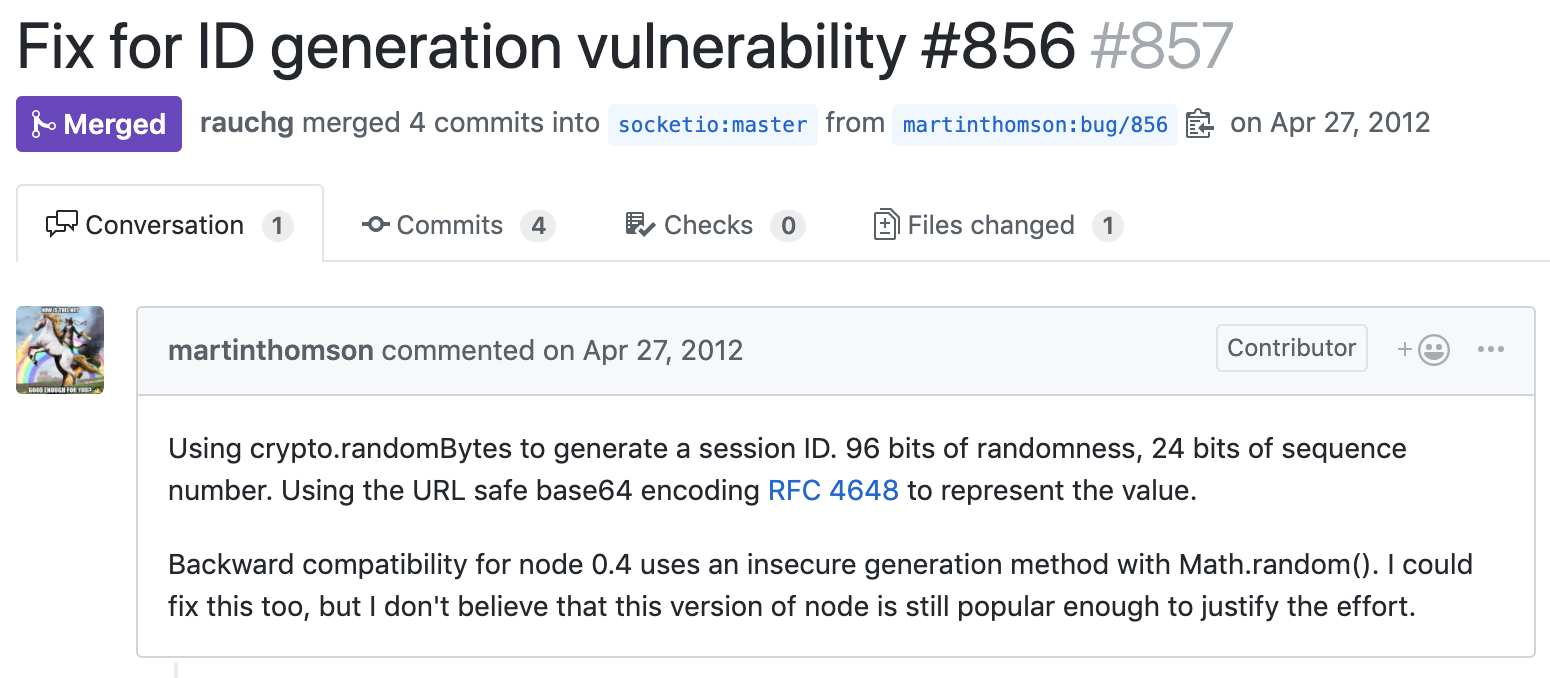}
        \label{fig:github_pull_request}
    }
    \subfigure[Fixing commit to mitigate the vulnerability.]
    {
        \includegraphics[width=1\linewidth]{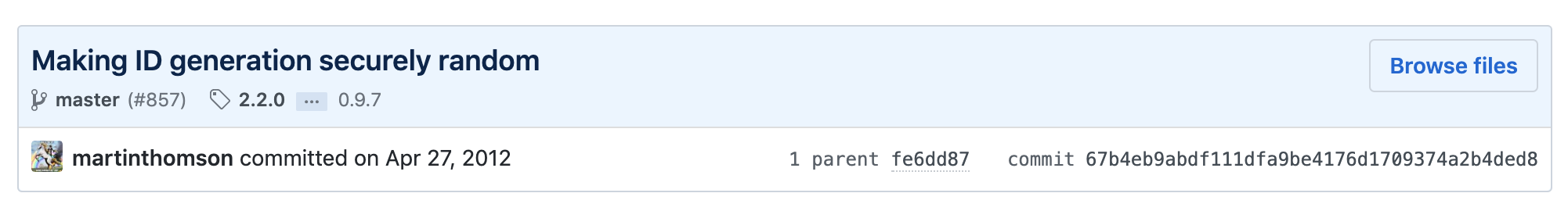}
        \label{fig:github_commit}
    }
    \subfigure[New \fixupdate~has been release.]{
        \includegraphics[width=0.7\linewidth]{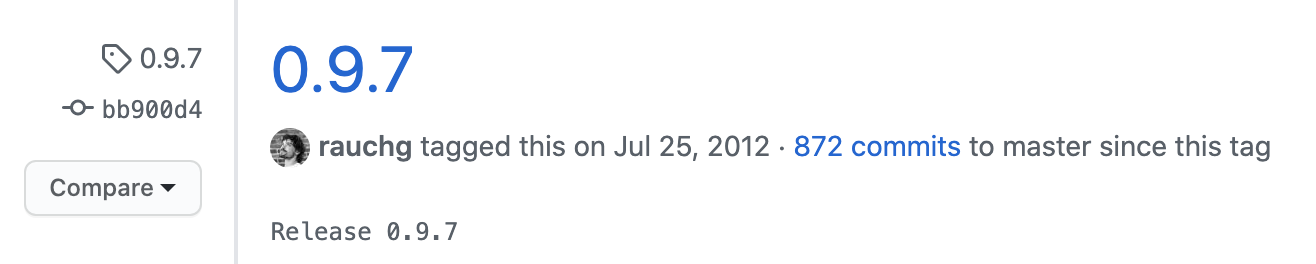}
        \label{fig:github_release}
    }
    \caption{Developer artifacts that mitigate a vulnerability (\texttt{socket.io}) on GitHub.}
    \label{fig:github_screenshot}
\end{figure*}

\subsection{Motivating Example}
\label{subsec:motivating_example}

Figure \ref{fig:github_screenshot} shows a practical case of the vulnerability fixing process which affects a library for network communication, i.e., \texttt{socket.io}.
Figure \ref{fig:vul_report} and Figure \ref{fig:github_issue} show step one, where the vulnerability is reported as a GitHub issue,\footnote{\url{https://github.com/socketio/socket.io/issues/856}} and summarized in \snyk.\footnote{\url{https://snyk.io/vuln/npm:socket.io:20120323}}
The vulnerability report contains detailed information regarding the identified problem, its severity, and a proof-of-concept to confirm the threat.
In this example, \texttt{socket.io} was vulnerable to a medium severity vulnerability.
We also found that the reporter is the same person who also created the fix.
Figure \ref{fig:github_pull_request} and \ref{fig:github_commit} show step two.
Figure \ref{fig:github_pull_request} shows a fix that will be merged into the code base.
The fix is submitted in the form of a pull request.\footnote{\url{https://github.com/socketio/socket.io/pull/857}}
Figure \ref{fig:github_commit} shows that there are four commits in a pull request with one commit actually fixes the vulnerability.\footnote{\url{https://github.com/socketio/socket.io/commit/67b4eb9abdf111dfa9be4176d1709374a2b4ded8}}
The other three commits were found to be unrelated, e.g., \texttt{"removing fixes for other bug"}.
Figure \ref{fig:github_release} shows that the \fixupdate~was available on July 25, 2012.\footnote{\url{https://github.com/socketio/socket.io/releases/tag/0.9.7}}

Interestingly, there is a \lag~in the vulnerability fix and release step. 
This example shows that there is an 89 days period between when the fix was created and released for any client to use.
We found that \texttt{socket.io} merged its fixes on April 27, 2012, however, it was actually released on July 25, 2012 and classified as a \fixupdatetypeS{patch}, i.e., V0.9.7. 
\section{Package-side Fix Commits and Landing: Preliminaries}
\label{sec:prelim_study}
From the motivating example in Section \ref{subsec:motivating_example}, we reveal how much development activity is directed towards fixing vulnerabilities compare to the other tasks.
Thus, we conduct a preliminary study to characterize \fixupdate~at the commit-level which including (i) changes of \fixupdate, (ii) contents of \fixupdate.
We first highlight the motivation, approach, and analysis to answer our preliminary questions.
We then show our data collection and finally provide the results.
The following two preliminary questions guide the study:

\paragraph{\textbf{\PqOneSen}}
\begin{itemize}
    \item \uline{\textit{Motivation}} 
    Our motivation for \PqOne~is to analyze the \fixupdate.
    Different from \citet{Decan:2018}, we manually investigate the version changes in the \fixupdate~itself.
    Our assumption is that \textit{every fix is applied as a \fixupdatetypeS{patch}}.
    
    \item \uline{\textit{Approach}}
    The approach to answer \PqOne~involves a manual investigation to identify the \fixupdate, i.e., \fixupdatetypeS{major}, \fixupdatetypeS{minor}, \fixupdatetypeS{patch}. 
    This is done in three steps.
    \begin{enumerate}
        \item The first step is to extract the fix-related information on GitHub.
        The extracted information is captured into three types as (i) an issue, (ii) a commit, and (iii) a pull request.
        \item The second step is to identify the release that contains a fix.
        This step involves an investigation of the package history.
        From the link in the first step, the first author manually tracked the git commit history to identify when the fix was applied.
        \item The final step is to identify a difference between a vulnerable release and a \fixupdate.
        We compare a vulnerable release, i.e., listed in the report, against a \fixupdate~to categorize the changes:
        (i) \fixupdatetypeS{major}, (ii) \fixupdatetypeS{minor}, and (iii) \fixupdatetypeS{patch}.
    \end{enumerate}
    \item \uline{\textit{Analysis}}
    The analysis to answer \PqOne~is the investigation of \fixupdate s.
    We use a summary statistic to show the \fixupdate~distribution.
    Furthermore, an interesting example case from the result is used to confirm and explain our findings.
\end{itemize}

\paragraph{\textbf{\PqTwoSen}}
\begin{itemize}
    \item \uline{\textit{Motivation}} 
    Extending \PqOne,
    our motivation for \PqTwo~is to reveal what kinds of contents are bundled within the \fixupdate.
    We complement recent studies, but at the commit-level \citep{Decan:2018, Hejderup2015, Decan:ICSME:2018}.
    We would like to evaluate the assumption that \textit{commits bundled in a \fixupdate~are mostly related to the \vulfix~commits}.
    \item \uline{\textit{Approach}} 
    The approach to answer \PqTwo~involves a manual investigation of contents inside the \fixupdate.
    This is done in three steps.
    \begin{enumerate}
        \item The first step is to gather information of a \vulfix~commit.
        In this case, we tracked the git commit history similar to the second step of \PqOne.
        \item The second step is to list commits in a \fixupdate.
        We use GitHub comparing changes tool to perform this task.\footnote{\url{https://help.github.com/en/github/committing-changes-to-your-project/comparing-commits}}
        \item The final step is to identify the type of commits bundled in a \fixupdate. 
        Similar to \PqOne, the first author manually tracked and labeled the commit as either (i) \vulfix~commit or (ii) other commit.
        To label commits, the first author uses source codes, commit messages and GitHub pull request information. 
        For validation, other co-authors confirmed the results, i.e., one author found the evidence and the other validated. 
    \end{enumerate}
 \item \uline{\textit{Analysis}}
    The analysis to answer \PqTwo~is to examine the portion of \vulfix~commits in a \fixupdate.
    We show the cumulative frequency distribution to describe the distribution of \vulfix~commits for 231 \fixupdate s.
    We use a box plot to show \vulfix~commit size in terms of lines of code (LoC).
    Similar to \PqOne, we show interesting example cases from the result.
\end{itemize}

\subsection{Data Collection}
\label{subsec:prelim_data_collection}

\begin{table}[]
\centering
\caption{A summary of package-side dataset information for preliminary study.}
\label{tab:prelim_dataset}
\scalebox{0.90}{
\begin{tabular}{@{}l|c@{}}
\toprule
\multicolumn{2}{c}{\textbf{npm Vulnerability Report Information}} \\ \midrule
Disclosures period & Apr 9, 2009 -- Aug 7, 2020 \\
\# vulnerability reports & 2,373 \\
\# samples of vulnerability reports (with fix references) & 231 \\ 
\# vulnerable packages & 172 \\\bottomrule
\end{tabular}
}
\end{table}

Our dataset contains the vulnerability reports with fix-related information on GitHub.
For the vulnerability reports, we crawled the data from \snyk~\citep{Web:snyk} that were originally disclosed in CVE and CWE database.
For the fix-related information, we focus on packages from the npm JavaScript ecosystem that is one of the largest package collections on GitHub \citep{Web:npm} and also has been the focus of recent studies \citep{Kikas:2017,Decan:2017,Abdalkareem:2017,Decan:2018,Hejderup2015,Decan:ICSME:2018}.
At the time of this study, we extracted fix-related information links directly from \snyk~(e.g., GitHub issue, commit, pull request).

As shown in Table \ref{tab:prelim_dataset}, we crawled and collected all reports from April 9, 2009 to August 7, 2020, i.e., in total 2,373 reports.
To identify the reports with fix-related information, we removed reports that (i) do not have the \vulfix~release or (ii) do not provide any fix-related information.
After that, we randomly select around 237 reports (10\% of 2,373) and manually filter reports that the vulnerable package does not follow semantic versioning.
In the end, the dataset for \PqOne~and \PqTwo~analysis consists of 231 reports that affect 172 packages.

\subsection{Results to the Preliminary Study}
\label{subsec:prelim_result}

\paragraph{{\textbf{\PqOneSen\newline}}}

\begin{table}[]
\caption{A summary statistic of \fixupdate~distribution in \PqOne.}
\label{tab:pq1_result}
\centering
\scalebox{0.90}{
\begin{tabular}{@{}lc@{}}
\toprule
\multicolumn{1}{c}{\textbf{\ffixupdate}} & \multicolumn{1}{c}{\textbf{\# of \vulfix~releases}} \\ \midrule
Major \fixupdatetype                               & 17 (7.36\%)                                                            \\
Minor \fixupdatetype                               & 47 (28.14\%)                                                            \\
Patch \fixupdatetype                               & \colorbox{red!25}{149 (64.50\%)}                                                            \\ \hline
&   231   \\ \bottomrule
\end{tabular}
}
\end{table}

Table \ref{tab:pq1_result} shows the evidence that not every fix is applied as a patch. 
This evidence contradicts our assumption.
We find that 64.50\% of fixes are a \fixupdatetypeS{patch}.
On the other hand, we find that 7.36\% and 28.14\% of fixes are \fixupdatetypeS{major}~and \fixupdatetypeS{minor}~respectively.
From our result, we suspect that some releases, especially \fixupdatetypeS{major}~and \fixupdatetypeS{minor}~might contain unrelated contents to the fixing commits.

The example case is a \fixupdatetypeS{major}~of an HTTP server framework, i.e., \texttt{connect} (V2.0.0). 
This package was vulnerable to Denial of Service (DoS) attack~\citep{Web:snyk:connect}. 
Under closer investigation, we manually validated that the other fixes were bundled in a \fixupdate, including API breaking changes (i.e., removed function).\footnote{\url{https://github.com/senchalabs/connect/blob/fe28026347c2653a9602240236fc43a8f0ff8e87/History.md\#200--2011-10-05}}
This fix also takes 53 days before it gets released.
We suspect that this may cause a \lag~in the \fixupdate, especially if the project has a release cycle.

\begin{tcolorbox}
\textbf{Summary:} We find that the vulnerability fix is not always released as its own patch update. 
We find that only 64.50\% of \vulfix~releases are a \fixupdatetypeS{patch}.
The rest of \vulfix~releases are either \fixupdatetypeS{major}, i.e., 7.36\%, or \fixupdatetypeS{minor}, i.e., 28.14\%.
\end{tcolorbox}

\paragraph{\textbf{\PqTwoSen\newline}}

\begin{figure*}[t!]
\centering
\includegraphics[width=.8\textwidth]{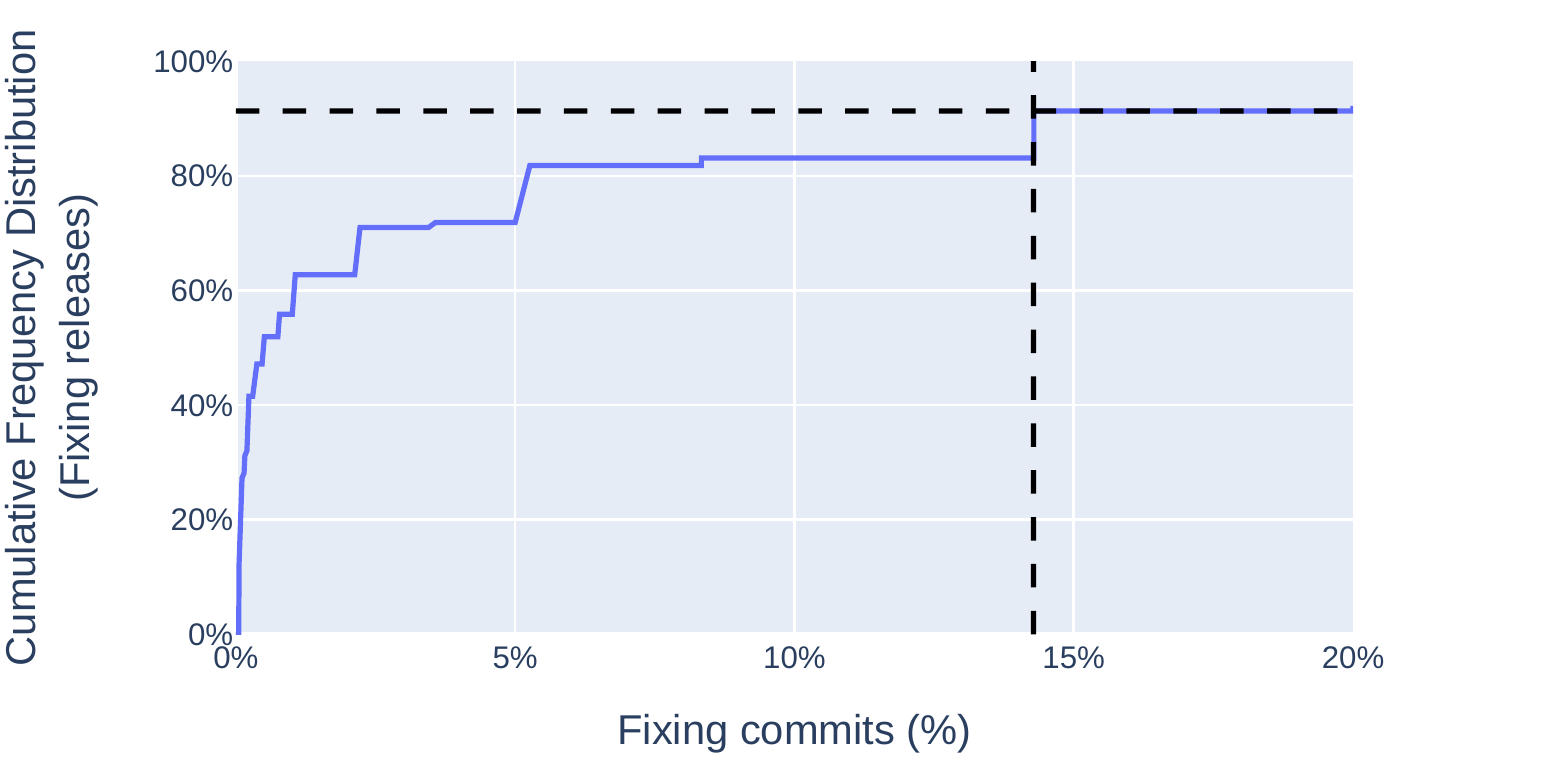}
\caption{
We find that 91.77\% out of 231 \vulfix~releases have \vulfix~commits up to 14.28\% of commits in a \fixupdate.
}
\label{fig:pq2_result}
\end{figure*}

\begin{figure*}[t!]
\centering
\includegraphics[width=.75\textwidth]{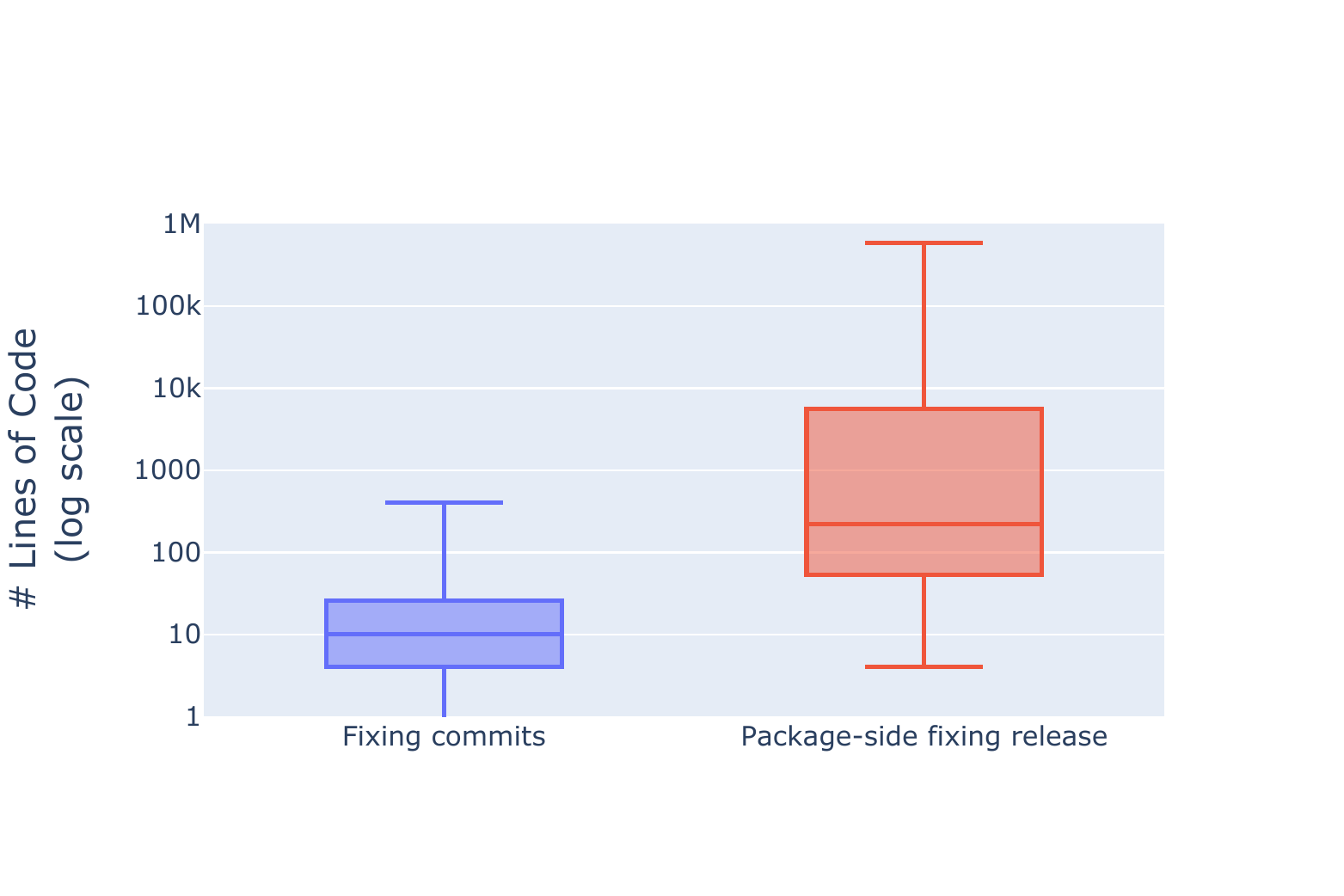}
\caption{LoC of the fixing commits for 231 vulnerabilities. We find that there are only few \vulfix~fix commits in the \fixupdate, i.e., median of 10.}
\label{fig:dist_fix_commit}
\end{figure*}

Figure \ref{fig:pq2_result} is evidence that fixes are usually bundled with other kinds of changes.
We find that 91.77\% out of 231 \vulfix~releases have up to 14.28\% commits that related to the fix, which means that 85.72\% of commits were unrelated.
Figure \ref{fig:dist_fix_commit} shows that the fix itself tends to contain only a few lines of code, i.e., median of 10 LoC.
Similar to the commit-level analysis, the \fixupdate~tends to contain a lot of changes, i.e., median of 219 LoC.
These results complement the finding of \PqOne~about \fixupdate~might contain unrelated changes to the fixing commit.

We show two examples to investigate the content of a fix and its size. 
The first example is a \fixupdatetypeS{patch}~of a simple publish-subscribe messing for a web, i.e., \texttt{faye} \citep{Web:snyk:faye}.
Under closer manual inspection, we find that there is one commit that updates the default value of variables.\footnote{\url{https://github.com/faye/faye/commit/e407e08c68dd885896552b59ce65503be85030ad}}
However, a \fixupdatetypeS{patch}~includes a total of 45 commits that is not related to the fix.\footnote{\url{https://github.com/faye/faye/compare/0.8.8...0.8.9}}
In the second example, we show that the actual fix is only a few lines of code. 
The \texttt{npm} package, which is the command line interface of a JavaScript package manager \citep{Web:snyk:npm}, took seven lines of code to fix the vulnerability.\footnote{\url{https://github.com/npm/npm/commit/f4d31693}}

\begin{tcolorbox}
\textbf{Summary:} We find that a small portion of the release contents related to the vulnerability fix, with 91.77\% of 231 \vulfix~releases have up to 85.72\% unrelated commits.
Furthermore, we found that the fix itself tends to have only a few lines of code (i.e., median of 10 lines of code).
\end{tcolorbox}
\section{Client-side Lags after the Package-side Fixes}
\label{sec:adoption_propagation}

The results of our preliminary study characterize the \fixupdate, where we find that (i) up to 64.50\% of vulnerability fixes are classified as a \fixupdatetypeS{patch}~and (ii) up to 85.72\% of commits in a release are unrelated to the actual fix.
Based on these results, we suspect that potential lags might occur while the \fixupdate~get adopted by the clients and transitively propagate throughout the dependency network.
Hence, we perform an empirical evaluation to explore potential \lags~in the adoption and propagation of the fix.

\subsection{Model and Track \llags}
\begin{figure}[]
    \centering
    \subfigure[An example of a \lag~in the propagation, caused by cascade delays from upstream clients.]
    {
        \includegraphics[width=.9\linewidth]{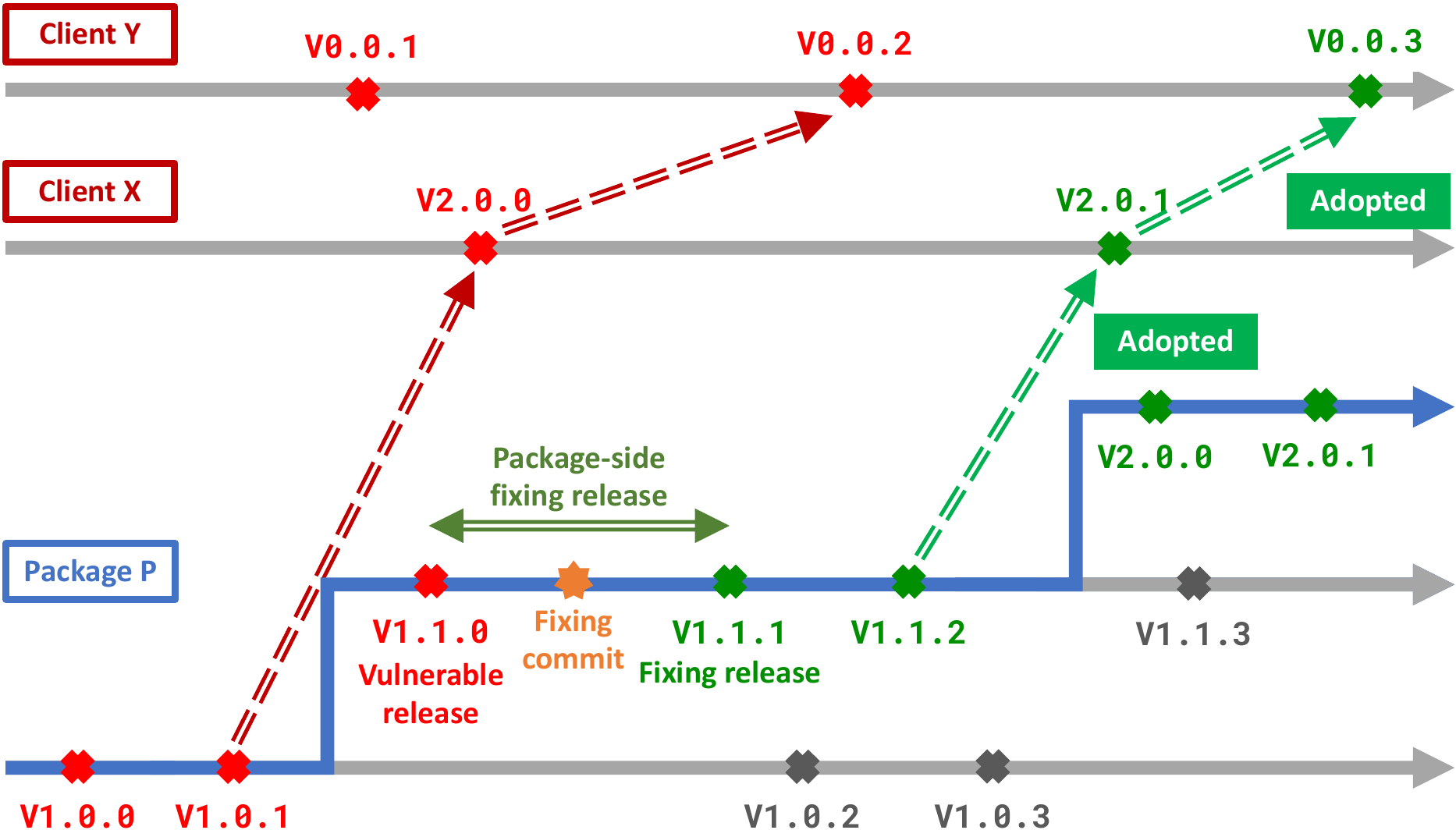}
        \label{fig:lag_in_adoption}
    }
    \subfigure[An example of a Latest Lineage (LL) and a Supported Lineage (SL) classify to track the freshness of a \fixupdate.]
    {
        \includegraphics[width=.9\linewidth]{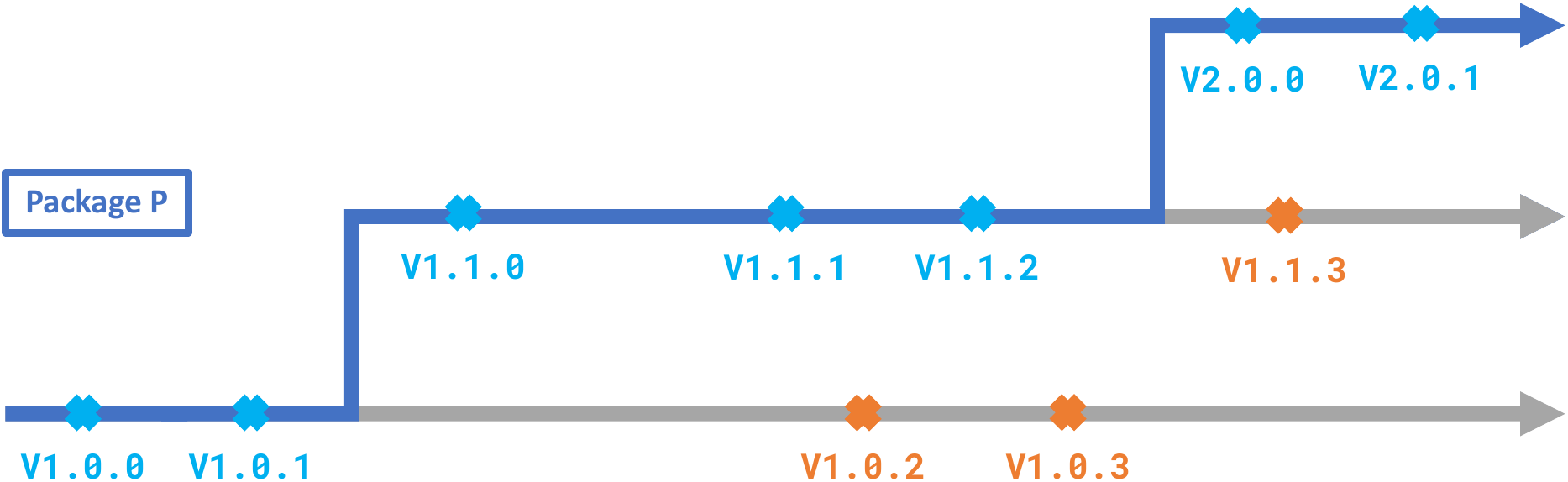}
        \label{fig:dependency_freshness}
    }
    \caption{These figures show the terms that are used to model and track the \lags.}
    \label{fig:measuring_lags}
\end{figure}

To explore potential \lags~in both adoption and propagation, we model and track the \fixupdate~and \clientupdate~as illustrated in Figure \ref{fig:measuring_lags}.

\paragraph{\uline{Released and Adopted by Version} - }
We identify \lags~in the adoption by analyzing the prevalence of patterns between a \fixupdate~and \clientupdate,
which is similar to technical lag \citet{Zerouali:ICSR2018} and based on semantic versioning.
The definition of \textit{\fixupdate} was explained in Section \ref{subsec:package-side_fixing_process} which describes how the package bumped the release version number.
Note that pre-releases or special releases are not considered in this study.
We then define a new term called a \textit{\clientupdate}.
\cclientupdate~describes how clients bumped the version of an adopted package up from vulnerable~version to \vulfix~release.
There are four kinds of \clientupdate:
(i) \texttt{\clientupdatetypeS{major}}~(major number of an adopted package is bumped up),
(ii) \texttt{\clientupdatetypeS{minor}}~(minor number of an adopted package is bumped up),
(iii) \texttt{\clientupdatetypeS{patch}}~(patch number of an adopted package is bumped up), and
(iv) \texttt{dependency removal}~(adopted package is removed from a client dependency list).

Figure \ref{fig:lag_in_adoption} shows an example of the two terms defined above.
First, we find that the \fixupdate~for package $\mathbb{P}$ is classified as a \fixupdatetypeS{patch}.
This is because of the difference between a \vulfix~release ($\mathbb{P}_{V1.1.1}$) and its vulnerable release ($\mathbb{P}_{V1.1.0}$).
Furthermore, we find that the \clientupdate~for client $\mathbb{X}$ is a \clientupdatetypeS{minor}.
This is because of the difference between the adopted \vulfix~release ($\mathbb{P}_{V1.1.2}$) and its previous vulnerable release ($\mathbb{P}_{V1.0.1}$).

\paragraph{\uline{Propagation Influencing Factors} - }
We define \textit{Hop} as the transitive dependency distance between a \fixupdate~and any downstream clients that have adopted this fix, i.e., one, two, three, and more than or equal to four hops.
As shown in Figure \ref{fig:lag_in_adoption}, client $\mathbb{X}$ is one hop away from package $\mathbb{P}$.
We consider two different factors to model and track \lags~in the propagation:

\begin{enumerate}
    \item \textit{Lineage Freshness:}
    refers to the freshness of the \fixupdate~as inspired by \citet{Cox-ICSE2015} and \citet{Raula:SANER2018}. 
    Figure \ref{fig:dependency_freshness} shows two types of lineage freshness based on the release branches including: 
    \textit{Latest Lineage (LL)}: the client has adopted any \fixupdate~on the latest branch, and \textit{Supported Lineage (SL)}: the client has adopted any \fixupdate~not on the latest branch. 
    Our assumption is that a \fixupdate~in the latest lineage is adopted faster than a \fixupdate~in a supported lineage, i.e., suffer less \lags.
    Figure \ref{fig:dependency_freshness} shows that three versions of package $\mathbb{P}$ (V1.0.2, V1.0.3, V1.1.3) are classified as SL. 
    
    \item \textit{Vulnerability Severity:}
    refers to the severity of vulnerability, i.e., H = high, M = medium, L = low, as indicated in the vulnerability report (as shown in Figure \ref{fig:vul_report} from Section \ref{sec:prelim}).
    Our assumption is that a \fixupdate~with higher severity is adopted quicker, i.e., less \lags.
\end{enumerate}

\subsection{Empirical Evaluation}
The goal of our empirical study is to investigate \lags~in the adoption and propagation.
We use these two research questions to guide our study:
\label{subsec:rq}
\paragraph{\textbf{\RqOneSen}}
Our motivation for \RqOne~is to understand whether developers are keeping up to date with the \fixupdate s.
We define that package-side and client-side fixing releases are consistent if \clientupdate~follow \fixupdate.
For example, \clientupdatetypeS{minor}~and \fixupdatetypeS{minor} combination is consistent, but \clientupdatetypeS{major}~and \fixupdatetypeS{patch} combination is not consistent.
Our key assumption is that the inconsistent combination requires more migration effort than the consistent one, which in turn is likely to create lags.

\paragraph{\textbf{\RqTwoSen}} 
Our motivation for \RqTwo~is to identify the existence of \lags~during a propagation. 
Concretely, we use our defined measures, i.e., propagation influencing factors, to characterize a propagation \lag.
Our assumption is that a \fixupdate~on the latest lineage with high severity should propagate quickly.

\paragraph{\uline{Data Collection} - }
\label{subsec:data_collection}
Our data collection consists of (i) vulnerability reports and (ii) the set of cloned npm package and client git repositories.
We use the same 2,373 vulnerability reports as shown in our preliminary study which crawled from \snyk~\citep{Web:snyk}.
As inspired by \citet{Wittern_MSR2016}, we cloned and extracted information of npm package and client from public GitHub repositories.
In this study, we consider only normal dependencies listed in the \texttt{package.json} file to make sure that the packages are used in the production environment.
Hence, other types of dependencies including: (1) devDependencies, (2) peerDependencies, (3) bundledDependencies, and (4) optionalDependencies are ignored in this study since they will not be installed in the downstream clients in the production or cannot be retrieved directly from the npm registry.
To perform the lags analysis, we first filter reports that do not have the \vulfix~release.
We then used the package name and its GitHub link from the reports to automatically match cloned repositories.

As shown in Table \ref{tab:dataset_information}, our data collection included 2,373 vulnerability reports that disclosed from April 9, 2009 to August 7, 2020.
There are 1,290 reports that already published the fixing releases which affect 786 different packages.
The statistics of vulnerable packages and reports are presented in the table.
For package and client repositories, we collected a repository snapshot from GitHub on August 9, 2020 with 152,074 repositories, 611,468 dependencies, and 1,553,325 releases.

\begin{table}[h]
\centering
\caption{A summary of the data collection which used to populate the dataset to answer \RqOne~and \RqTwo.}
\label{tab:dataset_information}
\scalebox{0.90}{
\begin{tabular}{@{}l|c@{}}
\toprule
\multicolumn{2}{c}{\textbf{npm JavaScript Ecosystem Information}} \\ \midrule
Repository snapshot & Aug 9, 2020 \\
\# package \& client GitHub repositories & 152,074 \\
\# total dependencies (with downstream) & 611,468 \\
\# total packages releases & 1,553,325 \\ \midrule
\multicolumn{2}{c}{\textbf{npm Vulnerability Report Information}} \\ \midrule
Disclosures period & Apr 9, 2009 -- Aug 7, 2020 \\
\# reports & 2,373 \\
\# reports (with fixing release) & 1,290 \\
\# vulnerable packages & 786 \\
\# vulnerabilities per package &  \\
\hspace{2em}- mean & 1.64 \\
\hspace{2em}- median & 1.00 \\
\hspace{2em}- SD & 2.58 \\
\# high severity vulnerabilities & 566 \\
\# medium severity vulnerabilities & 647 \\
\# low severity vulnerabilities & \hspace{1ex}77 \\ \bottomrule
\end{tabular}
}
\end{table}

\begin{table}[]
\centering
\caption{A summary of dataset information for the empirical study to answer \RqOne~and \RqTwo.}
\label{tab:rq_dataset_information}
\scalebox{0.90}{
\begin{tabular}{@{}l|c@{}}
\toprule
\multicolumn{2}{c}{\textbf{\RqOne~Dataset}} \\ \midrule
\# vulnerability reports (follow semver) & 410 \\
\# vulnerable packages (follow semver) & 230 \\
\# direct clients (follow semver) & 5,417 \\
\# filtered clients (break semver) & 4,000 \\ \midrule
\multicolumn{2}{c}{\textbf{\RqTwo~Dataset}} \\ \midrule
\# vulnerability reports (with fix released) & 617 \\
\# vulnerable packages (with fix released) & 344 \\
\# downstream clients & 416,582 \\ \bottomrule
\end{tabular}
}
\end{table}

\paragraph{\uline{Approach to Answer RQ1} - }
The data processing to answer \RqOne~involves the \fixupdate~and \clientupdate~extraction.
Similar to \PqOne, we first identify the \fixupdate~by comparing a vulnerable release and a 
\vulfix~release.
To track the \clientupdate, we then extract the direct clients' version history of the vulnerable packages.
A client is deemed vulnerable if its lower-bound dependency falls within the reported upper-bound as listed in a vulnerability report.

\begin{table}[]
\centering
\caption{A summary number of filtered clients grouped by their update pattern in \RqOne.
There are 4,000 packages and clients that excluded in the \RqOne.
}
\label{tab:rq1_filtering_result}
\scalebox{0.90}{
\begin{tabular}{@{}l|r@{}}
\toprule
\textbf{semver Update Patterns} & \textbf{\# clients} \\ \midrule
Major only & 94 \\
Minor only & 293 \\
Patch only & 2,812 \\
No change & 801 \\ \midrule
Filtered packages \& clients & 4,000 \\\bottomrule
\end{tabular}
}
\end{table}

To ensure quality, we additionally filter out packages and clients that did not follow semantic versioning as shown Table \ref{tab:rq1_filtering_result}. 
Our key assumption is to keep packages and clients that follow a semantic version release cycle, i.e., packages and clients should have all the update patterns of major landing, minor landing, and patch landing.
As a result, 4,000 packages and clients were filtered out from the dataset.
As shown in Table \ref{tab:rq_dataset_information}, our final dataset for \RqOne~consists of 410 vulnerability reports that affect 230 vulnerable packages and 5,417 direct clients.

The analysis to answer \RqOne~is the identification of \lags~in the adoption.
We show the frequency distribution of \clientupdate~in each \fixupdate.
In order to statistically validate our results, we apply Pearson's chi-squared test ($\chi^2$) \citep{Pearson:1900} with the null hypothesis \textit{`the \fixupdate~and the \clientupdate~are independent'}.
To show the power of differences between each \fixupdate~and \clientupdate~combination, we investigate the effect size using Cram\'er's V ($\phi'$), which is a measure of association between two nominal categories \citep{Cramer:1946}.
According to \citet{Cohen:1988}, since the contingency Table 6 has 2 degrees of freedom (df*), effect size is analyzed as follows: (1) $\phi'$~\textless~0.07 as Negligible, (2) 0.07~$\le$~$\phi'$~\textless~0.20 as Small, (3) 0.20~$\le$ $\phi'$~\textless~0.35 as Medium, or (4) 0.35~$\geq$~$\phi'$~as Large.
To analyze Cram\'er's V, we use the \texttt{researchpy} package.\footnote{\url{https://pypi.org/project/researchpy/}}

\paragraph{\uline{Approach to Answer RQ2} - }
The data processing to answer \RqTwo~involves propagation influencing factors extraction.
There are three steps to track downstream clients and classify lineage freshness and severity.
First, we build and traverse in a dependency tree for each \fixupdate~using a breadth-first search (BFS) approach.
The meta-data is collected from each downstream client which includes: (i) version, (ii) release date, and (iii) dependency list, i.e., exact version and ranged version.
We then classify whether or not a client is vulnerable using an approach similar to \RqOne.
Our method involves removing duplicated clients in the dependency tree, which is caused by the npm tree structure.
Second, we classify the lineage freshness of a \vulfix~release by confirming that it is on the latest branch.
Finally, we extract the vulnerability severity from the report.
As shown in Table \ref{tab:rq_dataset_information}, our final dataset for \RqTwo~consists of 617 vulnerability reports, 344 vulnerable packages with \vulfix~releases, and 416,582 downstream clients.

The analysis to answer \RqTwo~is the identification of \lags~in the propagation.
We show a summary statistic of \lags~in terms of days, i.e., the mean, the median, the standard deviation, and the frequency distribution, with two influencing factors.
In order to statistically validate the differences in the results, we apply Kruskal-Wallis non-parametric statistical test \citep{Kruskal:stat}.
This is a one-tailed test.\footnote{\url{https://github.com/scipy/scipy/issues/12231\#issuecomment-660404413}}
We test the null hypothesis that \textit{`\lags~in the latest and supported lineages are the same'}.
We investigate the effect size using Cliff's~$\delta$, which is a non-parametric effect size measure~\citep{Romano_SAIR2006}.
Effect size are analyzed as follows: (1) $|\delta|$~\textless~0.147 as Negligible, (2) 0.147~$\le$~$|\delta|$~\textless~0.33 as Small, (3) 0.33~$\le$ $|\delta|$~\textless~0.474 as Medium, or (4) 0.474~$\leq$~$|\delta|$~as Large.
To analyze Cliff's~$\delta$, we use the \texttt{cliffsDelta} package.\footnote{\url{https://github.com/neilernst/cliffsDelta}}
\subsection{Results to the Empirical Study}
\label{subsec:result}

\begin{table}[t]
\centering
\caption{A contingency table shows the frequency distribution of \clientupdate~for each \fixupdate.
We find that (i) there is a dependency between \fixupdate~and \clientupdate~and (ii) only the case of \fixupdatetypeS{minor} is consistent.}
\label{tab:rq1_contingency}
\scalebox{0.80}{
\begin{tabular}{@{}ccccr@{}}
\toprule
 & \ffixupdatetypeS{major}~(66) & \ffixupdatetypeS{minor}~(99) & \colorbox{red!25}{\ffixupdatetypeS{patch}~(245)} & \multicolumn{1}{c}{\textbf{All}} \\ \midrule
\cclientupdatetypeS{major} & \colorbox{green!25}{448 (46.18\%)} & 453 (30.00\%) & ~~657 (22.37\%) & 1,558 \\
\cclientupdatetypeS{minor} & ~~2 (0.21\%) & \colorbox{green!25}{761 (50.40\%)} & \colorbox{red!25}{1,082 (36.84\%)} & 1,845 \\
\cclientupdatetypeS{patch} & ~~0 (0.00\%) & ~~0 (0.00\%) & \colorbox{green!25}{~~625 (21.28\%)} & 625 \\
Dependency removal & \colorbox{yellow!25}{520 (53.61\%)} & 296 (19.60\%) & ~~573 (19.51\%) & 1,389 \\ \midrule
\textbf{All} & 970 & 1,510 & 2,937 & 5,417 \\ \bottomrule
\end{tabular}
}
\end{table}

\paragraph{\textbf{\RqOneSen}}
Our results are summarized into two findings.
First, Table \ref{tab:rq1_contingency} shows the evidence that most of \fixupdate s are \fixupdatetypeS{patch}s. 
As shown in the first row of a table, we find that 245 out of 410 \vulfix~releases have \fixupdatetypeS{patch}s~(highlighted in red).
We also find that there are 66 \fixupdatetypeS{major}s and 99 \fixupdatetypeS{minor}s.
This finding complements the result of \PqOne.

Second, Table \ref{tab:rq1_contingency} shows the evidence that there is a dependency between \fixupdate~and \clientupdate~variables.
However, there is no consistency across \fixupdate~and \clientupdate.
As highlighted in \textit{\cclientupdatetypeS{patch}} row of Table \ref{tab:rq1_contingency}, we find that there are only 21.28\% of clients adopt a \fixupdatetypeS{patch}~as \clientupdatetypeS{patch}s.
Instead, clients are more likely have \clientupdatetypeS{minor}s, i.e., 36.84\% of clients (highlighted in red).
For the case of \fixupdatetypeS{major}, there are 53.61\% of clients remove their dependencies to avoid vulnerability (highlighted in yellow).
The majority of clients that still adopt the \fixupdatetypeS{major}~are around 43.18\% as \clientupdatetypeS{major}.
The only case that we find consistent is \fixupdatetypeS{minor} which 50.40\% of clients adopt the fix as \clientupdatetypeS{minor} (highlighted in green).

For the statistical evaluation, we find that there is an association between the \fixupdate~and the \clientupdate.
Table \ref{tab:rq1_stat_test_result} shows that our null hypothesis on \textit{`the \fixupdate~and the \clientupdate~are independent'} is rejected (i.e., $\chi^2 = 1,484.48$, \textit{p-value} $< 0.001$).
From the Cram\'er's V effect size ($\phi'$), we got a value of 0.37 which shows the large level of association.

\begin{table}[]
\centering
\caption{A result of statistical test for \RqOne.
We find that differences between each distribution are significant and have a large level of effect size.}
\label{tab:rq1_stat_test_result}
\scalebox{0.90}{
\begin{tabular}{@{}ll@{}}
\toprule
\multicolumn{1}{c}{\textbf{Statistic}} & \multicolumn{1}{c}{\textbf{Value}} \\ \midrule
Pearson's chi-squared test ($\chi^2$) & $1,484.48$ \\
p-value & $< 0.001$ \\
Cram\'er's V ($\phi'$) & $0.37$ \\ \bottomrule
\end{tabular}
}
\end{table}

\begin{tcolorbox}
\textbf{Answer to \RqOne:} No for the case of \fixupdatetypeS{patch}, but yes for the others. 
We find that only 21.28\% of \fixupdatetypeS{patch}~are adopted as a \clientupdatetypeS{patch}~in clients. 
Instead, 36.84\% of \fixupdatetypeS{patch}~are adopted as a \clientupdatetypeS{minor}.
The evidence suggests that since clients keep stale dependencies, more migration effort is required to fix that client due to the potential risk from backward incompatible changes (i.e., \clientupdatetypeS{major}~or \clientupdatetypeS{minor}).
\end{tcolorbox}

\begin{table}[h]
\centering
\caption{A summary statistic of \lags~in the propagation (\# days) categorized by lineage freshness to show the difference between lags in LL and SL.
\llags~in the table is not accumulative.}
\label{tab:rq2_freshness_lags}
\scalebox{0.90}{
\begin{tabular}{@{}lcrrrr@{}}
\toprule
\multicolumn{1}{c}{\textbf{}} & \textbf{\# hop} & \multicolumn{1}{c}{\textbf{\# clients}} & \multicolumn{1}{c}{\textbf{Mean}} & \multicolumn{1}{c}{\textbf{Median}} & \multicolumn{1}{c}{\textbf{SD}} \\ \midrule
\multirow{4}{*}{LL} & 1 & 18,444 & 311.69 & \colorbox{red!25}{164.00} & 370.92 \\
 & 2 & 55,044 & 299.70 & \colorbox{red!25}{157.00} & 359.80 \\
 & 3 & 74,257 & 255.21 & \colorbox{red!25}{130.00} & 313.86 \\
 & $\geq$ 4 & 239,128 & 212.11 & \colorbox{red!25}{112.00} & 270.62 \\ \midrule
 & & 386,873 & & & \\ \midrule 
\multirow{4}{*}{SL} & 1 & 2,880 & 217.90 & \colorbox{green!25}{89.00} & 298.30 \\
 & 2 & 5,675 & 269.55 & \colorbox{green!25}{139.00} & 318.41 \\
 & 3 & 5,948 & 216.43 & \colorbox{green!25}{101.00} & 276.27 \\
 & $\geq$ 4 & 15,206 & 181.55 & \colorbox{green!25}{96.00} & 240.11 \\ \midrule
 & & 29,709 & & & \\ \bottomrule
\end{tabular}
}
\end{table}

\begin{table}[h]
\centering
\caption{A summary statistic of \lags~in the propagation (\# days) categorized by vulnerability severity to show the difference of lags between high, medium, and low severity vulnerability fixes.
\llags~in the table is not accumulative.}
\label{tab:rq2_serverity_survivability_stat}
\scalebox{0.90}{
\begin{tabular}{@{}lcrrrr@{}}
\toprule
\multicolumn{1}{c}{\textbf{}} & \textbf{\# hop} & \multicolumn{1}{c}{\textbf{\# clients}} & \multicolumn{1}{c}{\textbf{Mean}} & \multicolumn{1}{c}{\textbf{Median}} & \multicolumn{1}{c}{\textbf{SD}} \\ \midrule
\multirow{4}{*}{H} & 1 & 5,569 & 187.09 & \colorbox{green!25}{91.00} & 239.72 \\
 & 2 & 16,160 & 212.04 & \colorbox{green!25}{103.00} & 277.36 \\
 & 3 & 18,444 & 190.99 & \colorbox{green!25}{88.00} & 255.29 \\
 & $\geq$ 4 & 34,189 & 184.91 & 94.00 & 242.13 \\ \midrule
 &  & 74,362 &  &  &  \\ \midrule
\multirow{4}{*}{M} & 1 & 14,320 & 350.54 & \colorbox{red!25}{194.00} & 399.91 \\
 & 2 & 39,758 & 341.29 & \colorbox{red!25}{193.00} & 386.66 \\
 & 3 & 55,625 & 280.18 & \colorbox{red!25}{151.00} & 331.87 \\
 & $\geq$ 4 & 210,571 & 215.75 & \colorbox{red!25}{116.00} & 274.23 \\ \midrule
 &  & 320,274 &  &  &  \\ \midrule
\multirow{4}{*}{L} & 1 & 1,435 & 219.39 & 123.00 & 248.54 \\
 & 2 & 4,801 & 214.78 & 127.00 & 246.28 \\
 & 3 & 6,136 & 184.29 & 98.00 & 223.92 \\
 & $\geq$ 4 & 9,574 & 180.46 & 94.00 & 234.38 \\ \midrule
 &  & 21,946 & \multicolumn{1}{l}{} & \multicolumn{1}{l}{} & \multicolumn{1}{l}{} \\ \bottomrule
\end{tabular}
}
\end{table}

\begin{table}[t]
\centering
\caption{A comparison of \lags~in the propagation between clients that adopt the latest lineage and supported lineage \vulfix~release, i.e., by the median.
We find that difference between each distribution is significant mostly in the case of medium severity.
The effect sizes of those differences are negligible and small level.
(LL: median of the latest lineage, SL: median of the supported lineage, *: p-value \textless~0.001).}
\label{tab:rq2_statistical_test}
\scalebox{1}{
\begin{tabular}{@{}cccc@{}}
\toprule
\multicolumn{1}{c}{\textbf{\# hop}} & \multicolumn{1}{c}{\textbf{H}} & \multicolumn{1}{c}{\textbf{M}} & \multicolumn{1}{c}{\textbf{L}} \\ \midrule
\multirow{2}{*}{1} & LL \textgreater~SL & LL \textgreater~SL * & LL \textgreater~SL \\
& negligible & small & negligible \\ \midrule
\multirow{2}{*}{2} & LL \textgreater~SL * & LL \textgreater~SL * & LL \textgreater~SL * \\
& negligible & negligible & small \\ \midrule
\multirow{2}{*}{3} & LL \textgreater~SL & LL \textgreater~SL * & LL \textgreater~SL \\
& negligible & negligible & negligible \\ \midrule
\multirow{2}{*}{$\geq$ 4} & LL \textgreater~SL & LL \textgreater~SL * & LL \textless~SL * \\
& negligible & negligible & negligible \\ \bottomrule
\end{tabular}
}
\end{table}

\paragraph{\textbf{\RqTwoSen}}\xspace
Our results are summarized into two findings.
First, Table \ref{tab:rq2_freshness_lags} shows the evidence that the lineage freshness influences \lags~in a propagation.
As highlighted in red, we find that LL has more \lags~than SL in terms of days for every hops, e.g., median of \lags~for the first hop: 164 days $>$ 89 days.

Second, Table \ref{tab:rq2_serverity_survivability_stat} shows the evidence that the vulnerability severity influences \lags~in a propagation.
As highlighted in green, we find that the high severity \vulfix~release has the least \lags~than others in every hop, e.g., the first hop: 91 days. 
We also find that the medium severity fix has the most \lags~than others as highlighted in red, e.g., the first hop: 194 days.

For the statistical evaluation, we find that \lags~in the latest and supported lineage showed to have a significant (p-value $< 0.001$), but negligible to small association.
Table \ref{tab:rq2_statistical_test} shows that our null hypothesis on whether \textit{`\lags~in the latest and supported lineages are the same'} is rejected, i.e., the first hop to the more than the fourth hop for medium severity, the second hop and more than the fourth hop for low severity; and the second hop for high severity.

\begin{tcolorbox}
\textbf{Answer to \RqTwo:} Yes, lineage freshness and severity influence \lags~in the propagation.
We find that \vulfix~releases that occur on the latest lineage and medium severity suffer the most \lags.
\end{tcolorbox}
\section{Discussion}
\label{sec:discussion}
\subsection{Lessons Learned}
This section discusses three main implications based on our results in \PqOne, \PqTwo, \RqOne, and \RqTwo. 
These are presented as lessons learned and could have implications for both practitioner and researcher.

\begin{enumerate}

    \item \uline{Release cycle matters.}
    According to the results of \PqTwo, \vulfix~commits are small with less than 14.28\% of \vulfix~commits in the \fixupdate.
    We suspect that vulnerability fix repackage is a cause for \lags.
    Hence, developers of the vulnerable packages are recommended to release fixes as soon as they have applied the fix, if not, they should highlight these fixes when bundling the fix.
    Additionally, from 10 randomly selected vulnerabilities, we found that discussions between developers did not include an explicit mention of the vulnerability, i.e., GitHub issue, commit, and pull request.
    Since developers bundled the fix with other updates, developers may have been unaware. 
    \textit{In summary, researchers should provide strategies for making the most efficient update via the release cycle. 
    For example, (i) releasing an emergency patch that does not introduce any new features for security fixes to maintain backward compatibility and
    (ii) providing a security support for an active version, i.e., long-term-release version.
    Furthermore, practitioners can upgrade security fixes as first class citizens, so that the vulnerability fix can travel quicker throughout the ecosystem.}

    \item \uline{Awareness is important}.
    According to \PqOne~and \RqOne, 64.50\% of fixes are a \fixupdatetypeS{patch}.
    However, clients are more likely to have \clientupdatetypeS{major}~and \clientupdatetypeS{minor}, i.e., 22.37\% and 36.84\%, than the patch \clientupdatetype, i.e., 21.28\%.
    Security fixes need to be highlighted in the update note, as a possible reason is for failure to update because client developers are more interested in major features that are highlighted in an update.
    Recently, some open source communities start to make tools to highlight the vulnerability problems in a software ecosystem.
    GitHub \citep{Web:github_security_noti, Web:github_dependabot} made a new function for notifying a new vulnerability from the dependency list of clients by using a bot.
    However, GitHub stated that the tool will not be able to catch everything and send the alert notification within a guaranteed time frame.
    Also, npm \citep{Web:npm_audit} made a new command for listing the vulnerability information in downloaded dependencies of clients and try to automatically fix them called \texttt{npm audit}.
    However, there are some cases that \texttt{npm audit} does not work.
    For example, the immediate dependency does not adopt the \fixupdate~from the vulnerable package \citep{Web:npm_audit_manual}.
    In these cases, a manual review is required.
    Thus, client developers have to wait for the propagation of the \vulfix~release, i.e., a lag exists.
    According to \RqOne, 1,389 of 5,417 clients, i.e., 25.64\%, decided to remove vulnerable dependencies to mitigate the risk of vulnerabilities.
    Instead of waiting for the vulnerability fix, clients might remove the vulnerable dependency if they are able to find a similar package as a replacement or do not want to use the vulnerable dependency anymore.
    
    From a result of \PqTwo, explicit \fixupdate~with a highlight of the vulnerability is needed to speed up the adoption.
    \textit{In summary, researchers and practitioners need to provide developers more awareness mechanisms to allow quicker planning of the update. The good news is current initiatives like GitHub security are trending towards this.}
    
    \item \uline{Migration cost effort.} 
    From our first finding of \RqTwo, the \fixupdate~in the latest lineage suffers more lags than the supported lineage in terms of days.
    A possible reason is that developers of clients consider a package in the supported lineage is more worthwhile to adopt the new release than the latest lineage regardless of the fix.
    In terms of security, according to the official documentation of npm \citep{Web:npm_responding_to_threats}, when a security threat is identified, the following severity policy is put into action: (a) P0: Drop everything and fix!, (b) P1: High severity, schedule work within 7 days, (c) P2: Medium severity, schedule work within 30 days, (d) P3: Low severity, fix within 180 days.
    Surprisingly, the second finding of \RqTwo~shows evidence that low severity fixes are adopted quicker than medium severity fixes.
    A possible reason for the quicker adopting of low severity could be because the fix is easier to integrate into the application.
    \textit{In summary, researchers and practitioners that are package developers in npm seem to require additional time before updating their dependencies. }

\end{enumerate}
\subsection{Threats to Validity}
\label{sec:threats_to_validity}

\textit{Internal Validity -}
We discuss three internal threats.
The first threat is the correctness of the tools and techniques used in this study.
We use the listed dependencies and version number as defined in the \texttt{package.json} meta-file.
The threat is that sometimes some dependencies are not listed or the semantic version is invalid and vice-versa, so we applied a filter to remove clients that do not follow the semantic versioning, thus making this threat minimal.
The second threat is the tools used to implement our defined terminology (i.e., 
\texttt{numpy}, \texttt{scipy}, \texttt{gitpython}, and \texttt{semantic-version}).
To mitigate this, we carefully confirmed our results by manually validating the results for \RqOne~and \RqTwo, then also manually validating results with statistics on the npm website.
For the existing tool for suggesting the \fixupdate~like \texttt{npm audit}, we found that this tool was inappropriate to use in this work due to its limitation of the \texttt{package-lock.json} file is required for analyzing repositories, which only 2.27\% of repositories of the dataset and the tool assumes the latest information.
Unlike our work, we analyzed the data available in the historical snapshot.
As the correctness of dependency relations depends on getting all dependencies, the final internal threat is the validity of our collected data. 
In this study, our ecosystem is made up of packages and clients that were either affected directly or transitively by at least a single vulnerability.
We also make sure that the packages and their repositories are actually listed on a npm registry.
We are confident that the results of \RqTwo~are not affected by invalid data.

\textit{External Validity -}
The main external threat is the generality of other results to other ecosystems.
In this study, we focused solely on the npm JavaScript ecosystem.
However, our analysis is applicable to other ecosystems that have similar package management systems, e.g., PyPI for Python, Maven for Java. 
Immediate future plans include studying the \lags~in other ecosystems.
Another threat is the sample size of the analyzed data.
In this study, we analyzed only 1,290 vulnerability reports with \fixupdate s from 2,373 extracted reports from snyk.io.
This small size of sample data might not be able to represent the population.
However, we are confident of the data quality and reduced bias as we followed strict methods to validate by two authors for \PqOne, \PqTwo, and \RqOne~data.

\textit{Construct Validity -}
The key threat is that there may be other factors apart from the two factors, i.e., lineage freshness and vulnerability severity.
These factors are based on prior studies, i.e., measuring of dependency freshness from \citet{Cox-ICSE2015}, exploring the impact of vulnerability to transitive dependencies from \citet{Kikas:2017}, responding to a vulnerability \citet{Web:npm_responding_to_threats}. 
For future work, we would like to investigate other factors.
\section{Related Work}
\label{sec:related_work}
Complementary related works are introduced throughout the paper, in this section, we discuss some key related works.

\textit{On Updating Dependencies -}
These studies relate to the migration of libraries to the latest versions of libraries.
With new libraries and newer versions of existing libraries continuously being released, managing a system library dependencies is a concern on its own. 
As outlined in \citet{RaemaekersICSM,Teyton2012,Bogart:2016}, dependency management includes making cost-benefit decisions related to keeping or updating dependencies on outdated libraries. 
Additionally, \citet{Robbes:2012,hora:2015,Sawant2016,Bavota:2015,Ihara_OSS2017} showed that updating libraries and their APIs are slow and lagging.
\citet{Decan:2017} showed the comparison of dependency evolution and issue from three different ecosystems.
Their results showed that these ecosystems faced the dependency update issue which causes a problem to downstream dependencies, however, there is no perfect solution for this issue.
\citet{Kula:2017} found that such update decisions are not only influenced by whether or not security vulnerabilities have been fixed and important features have been improved, but also by the amount of work required to accommodate changes in the API of a newer library version.
\citet{Decan:ICSME:2018} performed an empirical study of technical lag in the npm dependency network and found that packages are suffered from the lags if the latest release of dependencies are not covered by their version ranges in \textit{package.json} file.
They also found that semantic versioning could be used in order to reduce the technical lag.
\citet{Mirhosseini:2017} studied about the pull request notification to update the dependencies.
Their results showed that pull request and badge notification can reduce lags, however, developers are often overwhelmed by lots of notifications.
Another study by \citet{Abdalkareem:2017} focused on the trivial packages and found that they are common and popular in the npm ecosystem.
They also suggested that developers should be careful about the selection of packages and how to keep them updated.
\textit{\textbf{Our work}} focuses on the \lag~of the \vulfix~release adoption and propagation with the influencing factors (i.e., lineage freshness and severity).
We also expand our study from prior works by taking transitive clients (i.e., downstream propagation) into consideration.
Our work complements the findings of prior work, with the similar goal of encouraging developers to update.

\textit{On Malware and Vulnerabilities -}
These studies relate to the security vulnerability within a software ecosystem from various aspects. 
\citet{Decan:2018} explored the impact of vulnerability within the npm ecosystem by analyzing the reaction time of developers from both vulnerable packages and their direct dependent packages to fix the vulnerabilities.
They also considered the reactions of developers from different levels of vulnerability severity.
They found that the vulnerabilities were prevalent and took several months to be fixed.
Several studies also explored the impact of vulnerability within various ecosystems by analyzing the dependency usage \citep{Kikas:2017,Linares:2017,Hejderup2015,Lauinger2016ThouSN}.
Some studies tried to characterize the vulnerability and its fix in various ecosystems other than the npm ecosystem \citep{Li:2017, Piantadosi:2019}.
There is a study about the relationship between bugs and vulnerabilities, to conclude that the relationship is weak \citep{Munaiah:2017}.
In order to increase the developers' awareness of the security vulnerability, some studies tried to create a tool to detect and alert vulnerability when it disclosed \citep{Cadariu:2015,Web:github_vul_view}.
There is a study about addressing the over-estimation problem for reporting the vulnerable dependencies in open source communities and the impact of vulnerable libraries usage in the industry \citep{Pashchenko:2018}.
Additionally, some studies tried to predict the vulnerability of software systems by analyzing the source code \citep{Shin:2008,Chowdhury:2011,Alhazmi:2007}.
\textit{\textbf{Our work}} takes a look at the \fixupdate~and its fix at the commit-level of npm package vulnerabilities, instead of the release-level in prior studies.
We also propose a set of definitions to characterize the package-side fixing release and client-side lags, which covers both direct and transitive clients.
Prior works, instead, only analyze the direct clients.

\textit{Mining-related Studies -}
These studies relate to the mining techniques in the software repository and software ecosystem.
The first step of software repository mining is data collection and extraction. Researchers need to have data sources and know about which part of the data can use in their work. 
In the case of the npm package repository, we can extract the information of packages from \texttt{package.json} meta-file \citep{Wittern_MSR2016, Mirhosseini:2017}.
In the case of the security vulnerability, we can collect data from Common Weakness Enumeration (CWE) \citep{Web:cwe} and Common Vulnerabilities and Exposures (CVE) \citep{Web:cve} database \citep{Linares:2017,Munaiah:2017,Lauinger2016ThouSN,Cadariu:2015,Alhazmi:2007,Chowdhury:2011}.
To study the issues within the software ecosystem, we also define the traversal of the downstream clients by using the dependency list of clients.
These studies introduce some techniques to model the dependency graph \citep{Kikas:2017,Hejderup2015,Bavota:2015}.
\textit{\textbf{Our work}} uses similar mining techniques to extract the dependencies as well as construct the ecosystem.
In our work, we manually extract and investigate the commits to understand the contents of the fix. 
\section{Conclusion}
\label{sec:conclusion}

Security vulnerability in third-party dependencies is a growing concern for software developers as the risk of it could be extended to the entire software ecosystem.
To ensure quick adoption and propagation of a \vulfix~release, we conduct an empirical investigation to identify \lags~that may occur between the vulnerable release and its \vulfix~release from a case study of npm JavaScript ecosystem.
We found that the \fixupdate~is rarely released on its own, with up to 85.72\% of the bundled commits in a \fixupdate~being unrelated to the fix.
We then found that a quick \fixupdate~(i.e., \fixupdatetypeS{patch}) does not always ensure that a client will adopt it quicker, with only 17.69\% of clients matching a \fixupdatetypeS{patch}~to a \clientupdatetypeS{patch}.
Furthermore, factors such as the lineage freshness and the vulnerability severity have a small effect on its propagation.

In addition to theses \lags~that we identified and characterized, this paper lays the groundwork for future research on how to mitigate these propagation \lags~in an ecosystem.
We suggest that researchers should provide strategies for making the most efficient update via the release cycle.
Practitioners also need more awareness to allow quicker planning of the update.
Potential future avenues for researchers include (i) a developer survey to a better understanding of the reason for releasing and adopting fixes, 
(ii) a performance improvement plan for highlighting the \vulfix~release tool, (iii) a tool for managing and prioritizing vulnerability fixing process.

\section*{Acknowledgment}

This work was supported by the Japan Society for Promotion of Science (JSPS) KAKENHI Grant Numbers JP18H04094, JP18H03221, 20K19774, 20H05706.

\bibliographystyle{spbasic}      %
\bibliography{bibliography.bib}   %

\begin{figure}[h]
  \includegraphics[width=1.75in,clip,keepaspectratio]{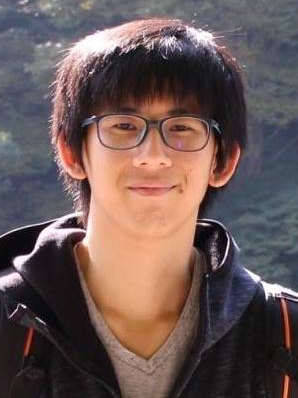}
\end{figure}
\textbf{Bodin Chinthanet} is a Ph.D. student at Nara Institute of Science and Technology, Japan. 
His research interests include empirical software engineering and mining software repositories. 
He is focusing on the security vulnerabilities in software ecosystems and how developers react to vulnerabilities in their software projects. 
Website: \url{https://bchinthanet.com/}.

\begin{figure}[h]
  \includegraphics[width=1.75in,clip,keepaspectratio]{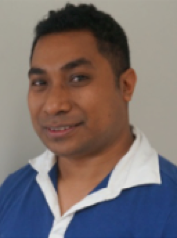}
\end{figure}
\textbf{Raula Gaikovina Kula} is an Assistant Professor at Nara Institute of Science and Technology. 
He received the Ph.D. degree from Nara Institute of Science and Technology in 2013. 
His interests include Software Libraries, Software Ecosystems, Code Reviews and Mining Software Repositories.

\begin{figure}[h]
  \includegraphics[width=1.75in,clip,keepaspectratio]{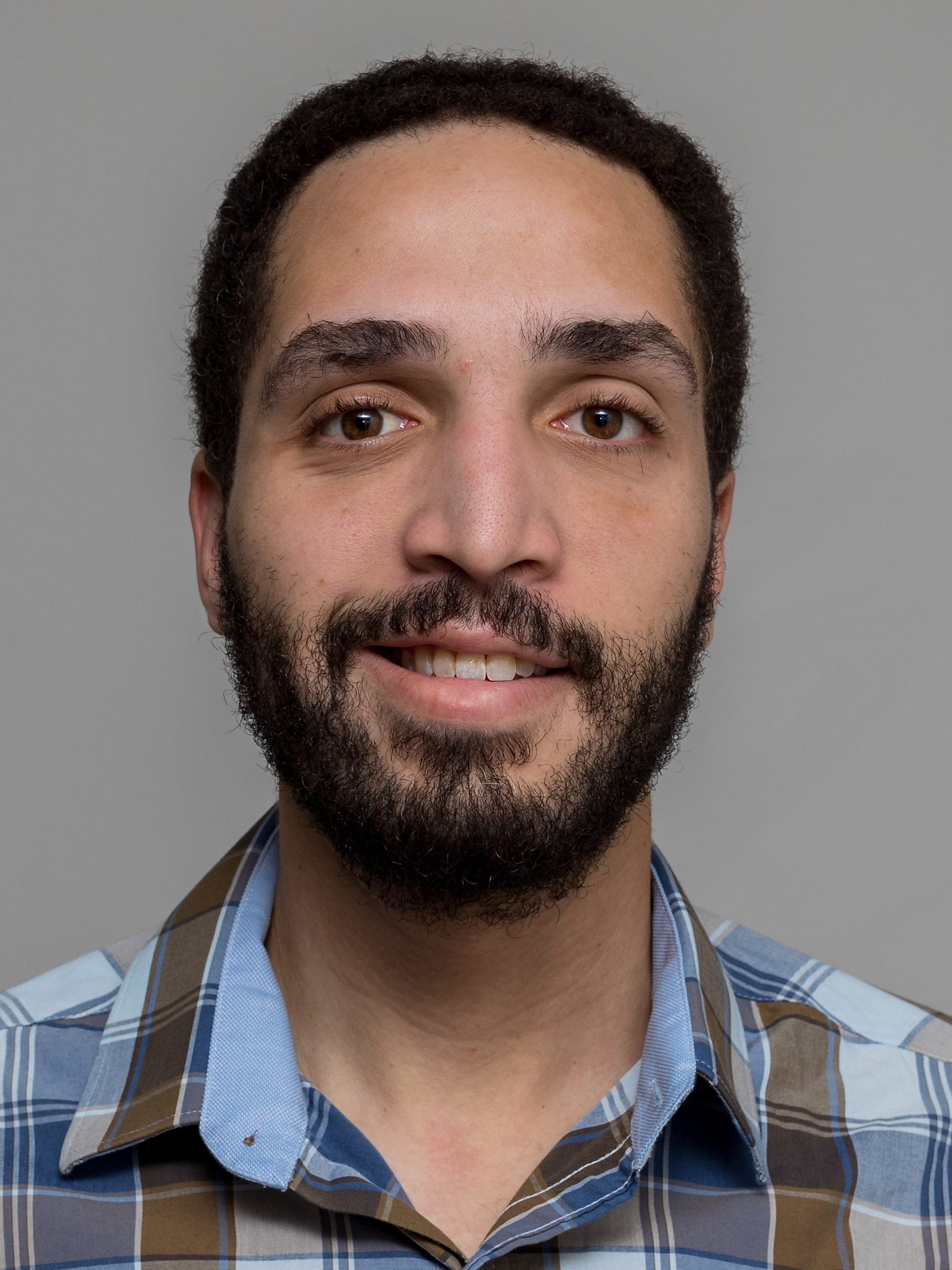}
\end{figure}
\textbf{Shane McIntosh} is an Associate Professor at the University of Waterloo. 
Previously, he was an Assistant Professor at McGill University. 
He received his Ph.D. from Queen’s University. 
In his research, Shane uses empirical methods to study build systems, release engineering, and software quality. 
Website: \url{http://shanemcintosh.org/}.

\begin{figure}[h]
  \includegraphics[width=1.75in,clip,keepaspectratio]{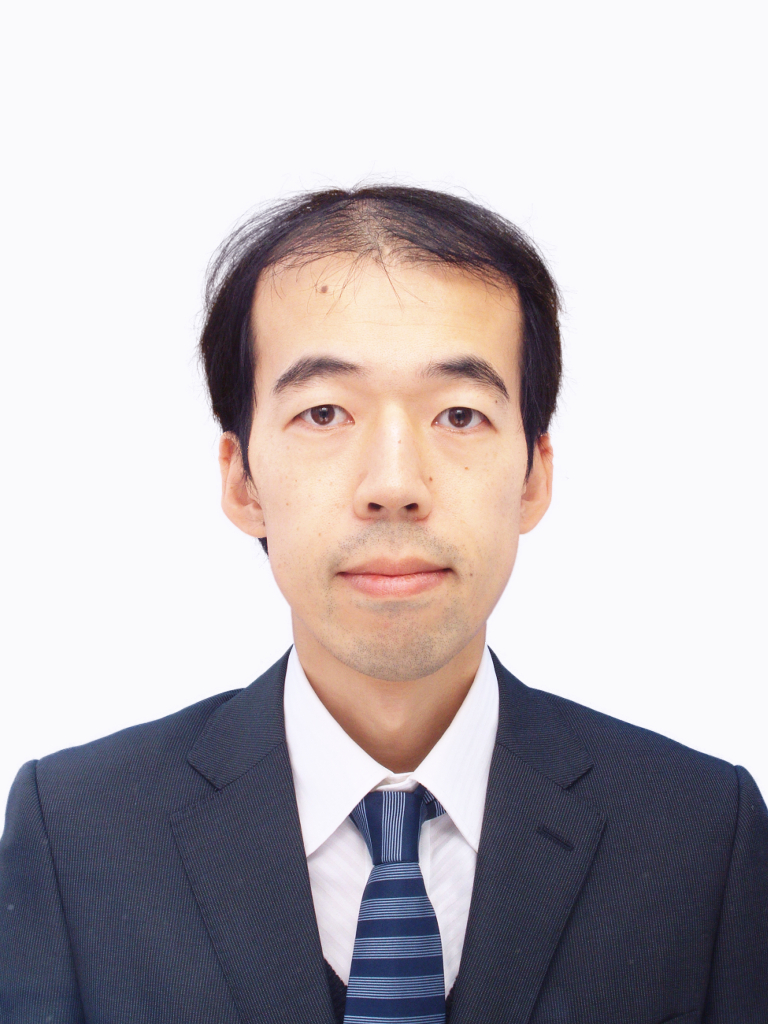}
\end{figure}
\textbf{Takashi Ishio} received the Ph.D. degree in information science and
technology from Osaka University in 2006.
He was a JSPS Research Fellow from 2006-2007.
He was an assistant professor at Osaka University from 2007-2017.
He is now an associate professor of Nara Institute of Science and
Technology.
His research interests include program analysis and software reuse.

\begin{figure}[h]
  \includegraphics[width=1.75in,clip,keepaspectratio]{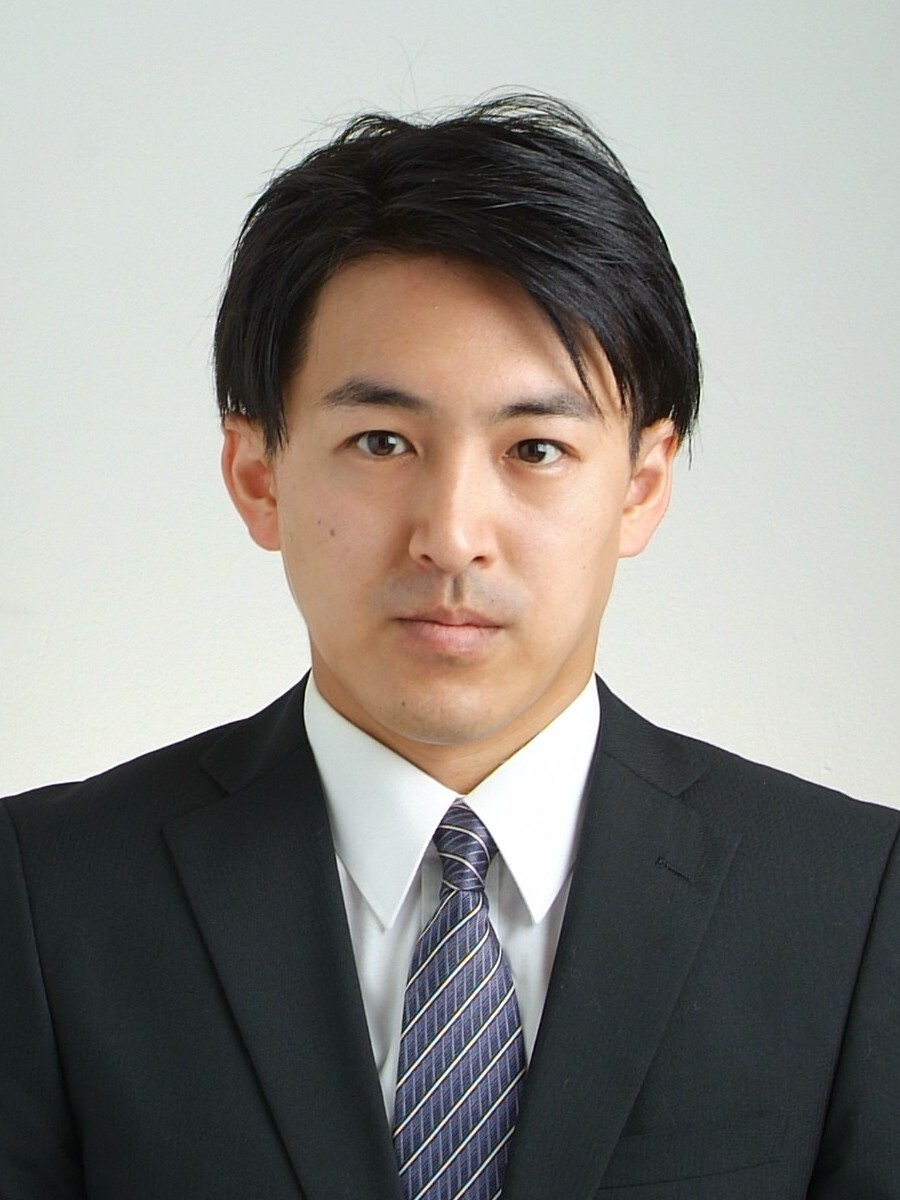}
\end{figure}
\textbf{Akinori Ihara} is a lecturer at Wakayama University in Japan. 
His research interests include empirical software engineering, open source software engineering, social software engineering and mining software repositories (MSR). 
His work has been published at premier venues like ICSE, MSR, and ISSRE. 
He received the M.E. degree (2009) and Ph.D. degree (2012) from Nara Institute of Science and Technology. 
More about Akinori and his work is available online at \url{http://www.wakayama-u.ac.jp/~ihara/}.

\begin{figure}[h]
  \includegraphics[width=1.75in,clip,keepaspectratio]{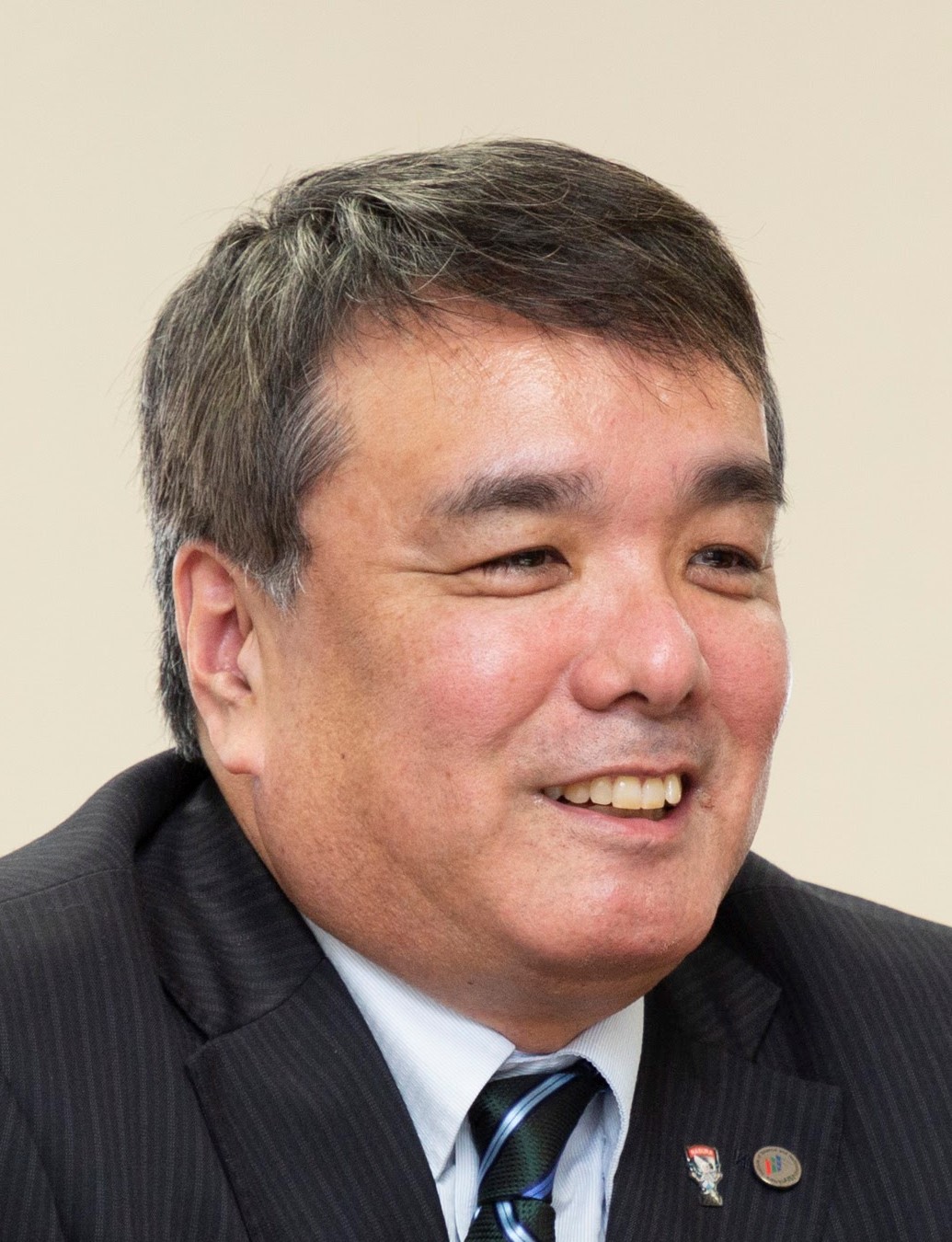}
\end{figure}
\textbf{Kenichi Matsumoto} is a professor in the Graduate School of Science and
Technology at Nara Institute Science and Technology, Japan. 
His research interests include software measurement and software process.
Matsumoto has a Ph.D. in information and computer sciences from Osaka University. 
He is a senior member of IEEE, and a member of the IEICE, and the IPSJ. 
Contact him at matumoto@is.naist.jp.

\end{document}